\documentclass{article}
\usepackage[margin=1in]{geometry}
\usepackage{graphicx} 
\usepackage{lineno}
\usepackage{todonotes}
\usepackage{empheq}
\usepackage{cancel}
\usepackage{amsmath}
\usepackage{amssymb}
\usepackage{hyperref}
\usepackage{tabularx}
\usepackage{ulem}

\title{Super-Gaussian approximations to optimum far-field irradiance in intersatellite optical communications: coherent and incoherent beam shaping}

\author{
Mario Badás Aldecocea$^{1,2}$,
Yves van Gennip$^{1}$,
Pierre Piron$^{1}$,\\
Jasper Bouwmeester$^{1}$,
Jérôme Loicq$^{1}$\\[1ex]
{\small $^{1}$\textit{Delft University of Technology, Delft, the Netherlands}}\\
{\small $^{2}$\textit{University of Cambridge, Cambridge, United Kingdom}}\\
}
\date{}

\begin{document}

\maketitle

\begin{abstract}
    Intersatellite optical communication links are strongly affected by transmitter pointing jitter, which stochastically modulates the received optical power and degrades communication performance. While adjusting the divergence of a conventional Gaussian beams can partially mitigate this effect, the optimum far-field irradiance distribution for maximizing link performance has not been formally derived. In this work, we introduce a variational formalism to determine the optimum far-field irradiance for an intersatellite link affected by pointing jitter. The flat-top beam is found to be the optimum beam shape for minimizing outage probability. In particular, it requires $\sim37,8\%$ of the power needed by a conventional Gaussian beam to achieve the same outage probability. However, the flat-top profile is discontinuous and physically unrealizable. To address this, we analyze a continuous family of super-Gaussian beam shapes that approximate the flat-top as the order increases. In addition, several coherent and incoherent beam shaping techniques are evaluated to assess their ability to reproduce the optimum irradiance distribution. The results show that these techniques can reduce the required transmitted power by up to $\sim50\%$ compared with conventional Gaussian beams.
\end{abstract}

\section{Introduction}
\begin{figure}[!htb]
    \centering
    \includegraphics[width=0.9\linewidth]{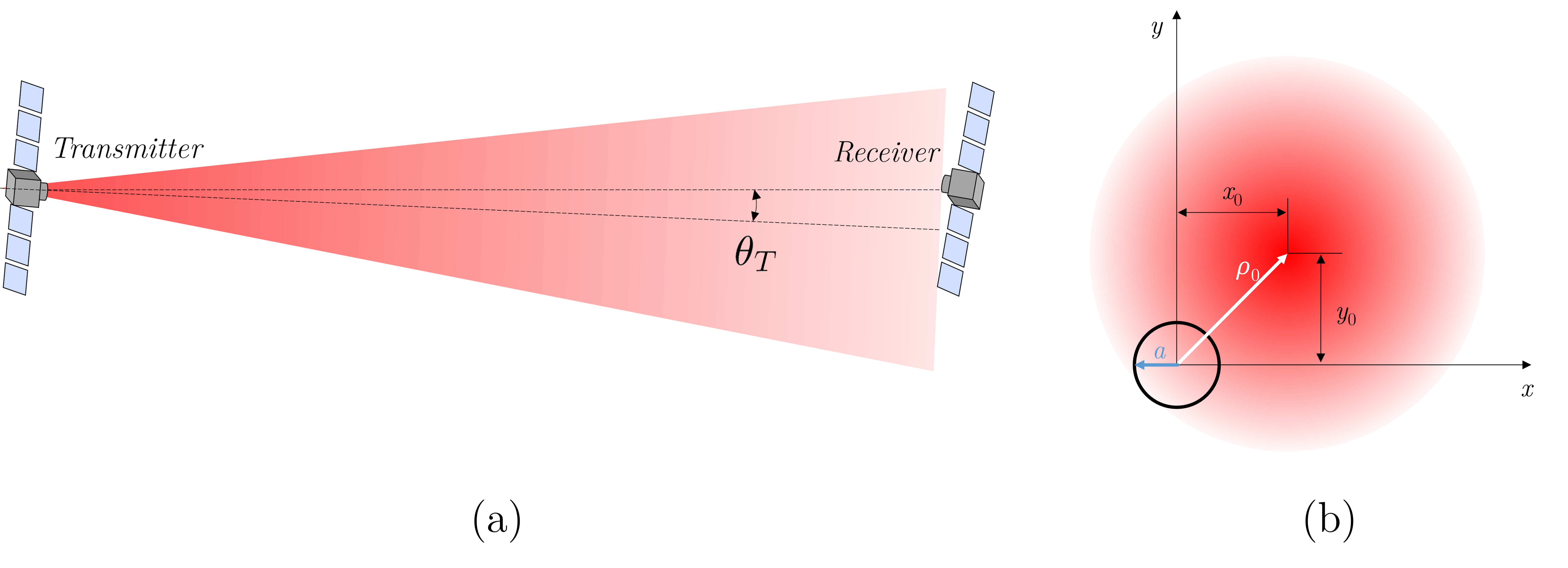}
    \caption{Effect of transmitter pointing jitter in intersatellite links (a) shows how an small angular deviation on the transmitted beam, deviates its center significantly in the receiver aperture plane and (b) shows the resulting deviation as seen by the receiver aperture plane, where $a$ is the radius of the receiver's aperture and $\rho_0$ is the instantaneous beam decenter.}
    \label{fig_pj}
\end{figure}

Intersatellite free-space optical communication (FSOC) links promise to form the backbone of future high-throughput global communication networks. Compared to classical radiofrequency links, FSOC offers higher bandwidth and narrower beam divergence, which has driven strong interest in this technology. However, satellite terminals operate in the harsh space environment, which includes thermal loads, radiation, and microvibrations \cite{badas_opto-thermo-mechanical_2023}. The latter is particularly critical for FSOC terminals, as it directly affects the pointing jitter of the transmitted laser beam (see Fig.~\ref{fig_pj}(a)).

To mitigate the impact of microvibrations, satellites employ onboard pointing mechanisms that typically reduce residual pointing jitter to a few microradians \cite{wang_angular_2021, riesing_pointing_2022}. Nevertheless, for the large distances involved in intersatellite links, this residual pointing jitter can displace the beam center at the receiver plane significantly with respect to the receiver aperture (see Fig.~\ref{fig_pj}(b)). Consequently, the stochastic nature of transmitter pointing jitter induces fluctuations in the power captured by the receiver aperture, which in turn impacts the overall communication performance.

A common strategy to address this issue is to adjust the divergence of a Gaussian beam in order to mitigate the effects of pointing jitter \cite{toyoshima_optimum_2002, farid_outage_2007}. More recently, studies have proposed generating optimized non-Gaussian beam profiles to further improve system performance \cite{badas_optimum_2024, badas_metalens_2025}. However, a formal framework for determining the optimum far-field irradiance shape that maximizes a given communication parameter is still lacking. In this paper, we introduce a variational formalism to derive the optimum far-field irradiance distribution for an intersatellite link affected by transmitter pointing jitter. Furthermore, several optical systems capable of approximating these optimum far-field irradiance profiles are analyzed.

First, the mathematical model of the intersatellite free-space channel is presented in \S~\ref{sec_channel}, where the power statistics and the communication performance of the optical link are derived. Next, \S~\ref{sec_variational_FP} proposes a variational method to obtain the optimum far-field irradiance. Since the resulting optimum far-field irradiance exhibits a discontinuity, \S~\ref{sec_smallap} analyzes a continuous family of super-Gaussian beam shapes that approximate this optimum field. Finally, in \S~\ref{sec_beamshaping} incoherent and coherent beam-shaping techniques are studied to approximate the optimum beam shapes identified in the previous sections.

\section{Mathematical model of the free space channel}\label{sec_channel}

Considering the pointing jitter problem stated above, we aim to determine the optimum far-field irradiance distribution $I(x,y)$ that maximizes the performance of an intersatellite optical link under the influence of transmitter pointing jitter. Formally, the problem can be stated as:

\textit{Find the far-field irradiance $I(x,y)\in \mathbb{R}$ with $(x,y)\in \mathbb{R}^2$, given a probability density function of the beam center location $f_\mathrm{XY}(x,y)\in\mathbb{R}$, that minimizes a chosen communication performance parameter $\mathcal S$.}

As the pointing jitter probability density function (PDF) will have azimuthal symmetry, the problem can be stated in terms of the radial coordinate $\rho$ instead. Typical communication performance metrics $\mathcal{S}$ to be minimized are: average bit error probability (ABEP) \cite{toyoshima_optimum_2002}, probability of fade, probability of outage and the inverse of the reliability (probability of being below a given bit error probability threshold) \cite{badas_seidel_2025,farid_link_2009}.
Using the variational formalism, minimizing any of these parameters is equivalent to finding an extremum of the functional $\mathcal{S}$ as
\begin{equation}\label{eq_variational}
\delta \mathcal{S} = 0
\end{equation}

For parameters such as outage, fade, and reliability, $\mathcal S$ can be expressed as the cumulative density of the received power below a threshold\footnote{The integration that defines $\mathcal{S}[I]$ is done over the interval $[0,P_\text{th})$, this is important in the formal mathematical sense, as including or not the extremes of the range considered can have a far-reaching conseuqences on the formalism.} $P_\text{th}$
\begin{equation*}
    \mathcal{S} [I] = \int_0^{P_\text{th}} \mathcal{F}_P\left[I, f_\text{XY}\right] dP
\end{equation*}
where $\mathcal{F}_P$ is the PDF of the received power, which depends on both the far-field irradiance $I(x,y)$ and the PDF of the pointing jitter $f_\text{XY}(x,y)$. 
The received power PDF, $\mathcal{F}_P$, can be obtained from the vector to scalar probability mapping of the beam center location to the collected power as \cite{rohatgi_introduction_2001}
\begin{equation}\label{eq_FPdef1}
    \mathcal{F}_P\left[I, f_\text{XY}\right] = \iint _{-\infty}^\infty f_{\text{XY}}(x_0,y_0)\;\delta \left[P-g(x_0,y_0)\right]\,dx_0 \; dy_0
\end{equation}
where $f_{\text{XY}}(x_0,y_0)$ is the probability density function of the location center of the beam in the far-field and $g(x_0,y_0)$ is the received power in the aperture when the center of the transmitted beam is located at $(x_0,y_0)$. Mathematically, this power can be written as
\begin{equation*}\begin{split}
    g(x_0,y_0)&= \iint_{\mathbb{R}^2} I(x,y)\,\mathcal{A}(x-x_0,y-y_0)\,dx\;dy
\end{split}\end{equation*}
where $\mathcal{A}(x,y)$ is the receiver aperture pupil function
\begin{equation*}
    \mathcal{A}(x,y)=\begin{cases}
        1\quad \text{if}\quad a<x^2+y^2 < b\\
        0\quad \text{if}\quad \text{else}
    \end{cases}
\end{equation*}
where $a$ and $b$ are the secondary and primary mirror radii of the receiver telescope. This integral is equivalent to a bidimensional convolution and can be expressed via the Fourier transform $\mathcal{F}$ as
\begin{equation}\label{eq_gx0y0}
     g(x_0,y_0) = \{ I\ast \mathcal{A}\} (x_0,y_0) 
    = \mathcal{F}^{-1}\{\mathcal{F}(I) \cdot\mathcal{F}(\mathcal{A})\}
\end{equation}

Therefore, the problem of finding the optimum far-field irradiance to maximize the performance of an intersatellite optical link impinged by transmitter pointing jitter has been converted into the variational problem given by Eqs.~\eqref{eq_variational}-\eqref{eq_gx0y0}.

When both the far field irradiance and the receiver aperture are azimuthally symmetric with respect to the optical axis, the power collected can be written as a function of the radial coordinate $g(x_0,y_0)\rightarrow g(\rho_0)$. Furthermore, for very large distances, the aperture is very small compared to the far-field irradiance footprint due to the divergence of the beam. Hence, the irradiance variation across the aperture is negligible in this regime, and the power as a function of the center of the beam can be written as
\begin{equation}\label{eq_grho0}
    g(\rho_0) \approx A\times I(\rho_0)
\end{equation}
where $A$ is the area of the receiver aperture. In general, the pointing jitter statistics will be azimuthally symmetric with respect to the optical axis. In this case, the pointing jitter statistics are given by the density function $f_\mathrm{R}(\rho)$. This condition, along with the azimuthal symmetry of the receiver aperture, will yield an axially symmetric far-field irradiance distribution $I(\rho)$. Specifically, the pointing jitter is usually considered a Rayleigh distribution in radial coordinates, resulting from a bivariate Gaussian distribution in both elevation and azimuthal angles \cite{badas_optimum_2024}. The Rayleigh density function for a scale parameter $\sigma$ (where $\sigma^2$ corresponds to the standard deviation of the underlying elevation and azimuth pointing jitter) is given by
\begin{equation}\label{eq_rayleigh}
    f_\mathrm{R}(\rho)=\dfrac{\rho}{2\pi\sigma^2}\exp\left(-\dfrac{\rho^2}{2\sigma^2}\right).
\end{equation}
Under these assumptions, the received power can be expressed as a function of the radial coordinate $P=g(\rho)$ and the received power statistics can be computed with the scalar to scalar $\mathbb{R}\rightarrow\mathbb{R}$ mapping of the density functions \cite{rohatgi_introduction_2001}
\begin{equation}\label{eq_FPdef2}
    \mathcal{F}_P(P) = \sum_{k=1}^{k_{\max}}\left|{\frac {d}{dP}}\rho_k\right|\cdot f_\mathrm{R}{\big (}\rho_k{\big )}  
\end{equation}
where $k_{\max}$ is the number of monotonical piecewise functions in between the extreme values of the function $\rho(P)$.
Furthermore, the condition of the conservation of energy needs to be included. This means, that if the beam has a total power $P_0$ in the transmitter aperture plane, the irradiance field at the receiver aperture plane $I(\rho)$ has to fulfill
\begin{equation}\label{eq_finitepower}
     P_0 =\iint_{\mathbb{R}^2}I(x,y) \;dx\;dy = 2\pi \int_{\mathbb{R}^+} I(\rho) \;\rho\;d\rho.
\end{equation}
\section{Optimum far-field irradiance via variational formalism}\label{sec_variational_FP}

Considering the variational statement given in the previous section, no closed solution for $I(x,y)$ has been found when imposing the functional $\mathcal{S}[I]$ to be minimized. Furthermore, exploiting the azimuthal symmetry to obtain $I(\rho)$ has also not provided any solution through the variational approach. However, applying both the subsidiary condition through Lagrange multipliers (for the finite power condition) and the Weierstrass-Erdmann corner conditions (for allowing discontinuities in the solution) as in Ref.~\cite{gelfand_calculus_2000}, has not proven successful. This is because the problem is ill-posed. As will be shown later, without enforcing all necessary conditions, the problem becomes degenerate, and no solution can be obtained. Incorporating these conditions within the aforementioned formalisms proved challenging; therefore, the problem was instead addressed by determining the optimal captured power probability density function, and only thereafter finding the irradiance profile that yields the optimum captured power probability density function.

Therefore, the optimization problem can be restated in terms of the probability density function of the received power, $\mathcal{F}_P(P)$. When doing so, if the optimum solution is found for the function $\mathcal{F}_P(P)$, then an irradiance distribution $I(\rho)$ needs to be found resulting in such a PDF. Mathematically, we want to make the functional stationary
\begin{equation}\label{eq_S}
\mathcal{S}[\mathcal{F}_P]
\;=\;
\int_0^{P_{\rm th}}\mathcal{F}_P(P)\,dP
\end{equation}
subject to the two constraints
\begin{align}
    &(1)\quad 0\leq P\leq P_{\max} \quad & \text{positivity and upper‐bound constraint}\\
    &(2)\quad \int_0^\infty\mathcal{F}_P(P)\;dP=1 & \text{normalization}\label{eq_norm}
\end{align}
The first implies that the total power of the beam is finite (through an upper limit of the achievable maximum power), while the second is the normalization condition of the probability density function. The $P_{\max}$ can be defined as
\begin{equation*}
    P_{\max}=\max_{(x_0,y_0)} g(x_0,y_0)
\end{equation*}
where $g(x_0,y_0)$ is the power collected when the center of the beam is at $(x_0,y_0)$ (see Eq.~\eqref{eq_gx0y0}).

To solve the variational problem, we first introduce a Lagrange multiplier \(\lambda\) to account for the normalization condition in Eq.~\eqref{eq_norm}. For that, we consider the augmented functional
\[
\Phi[\mathcal{F}_P]
\;=\;
\int_0^{P_{\rm th}}\mathcal{F}_P(P)\,dP
\;-\;
\lambda\Bigl(\int_0^{P_{\max}}\mathcal{F}_P(P)\,dP-1\Bigr).
\]
To include the stationarity condition ($\delta \Phi=0$), we take a small variation \(F_P(P)\mapsto F_P(P)+\varepsilon\,\eta(P)\) with an arbitrary test function \(\eta(P)\).  The first variation must vanish as
\[
\delta\Phi
=\int_0^{P_{\rm th}}\eta(P)\,dP
\;-\;
\lambda\int_0^{P_{\max}}\eta(P)\,dP
\;=\;0\quad\quad\forall\,\eta(\cdot).
\]

By splitting the last integral, the stationary condition given by
\[
\int_0^{P_{\rm th}} (1-\lambda)\;\eta(P)\,dP
\;-\;\int_{P_{\rm th}}^{P_{\max}}\lambda\;\eta(P)\,dP
=0 \quad\quad\forall\,\eta(\cdot).
\]
must hold. For this condition to be met for any arbitrary test function $\eta(P)$, the coefficients of \(\eta\) on the two intervals must be zero. This gives the pointwise {Euler–Lagrange} conditions
\begin{align*}
    & \lambda=1\quad \quad\text{for}\quad  0\le P\le P_{\rm th}\\
    & \lambda=0\quad\quad \text{for}\quad  P_{\rm th}< P\le P_{\max}
\end{align*}
These two requirements cannot both hold simultaneously as the multiplier $\lambda$ is a scalar. Hence, there is no smooth, everywhere‐positive density $\mathcal{F}_P$ satisfying the stationarity conditions. What survives are the admissible \textit{extremal} measures which lump all probability at a point.  In fact, the only way to satisfy normalization and make \(\delta\Phi=0\) is to place all mass at one of the boundaries of the integration ranges
\begin{align*}
    \text{Mass at}\;&P=0\rightarrow \quad&\mathcal{F}_P(P)&=\delta(P)\rightarrow \quad&\mathcal{S}=1\\
    \text{Mass at}\;&P=P_{\text{th}}\rightarrow \quad&\mathcal{F}_P(P)&=\delta(P-P_{\text{th}})\rightarrow\quad&\mathcal{S}=1/2\\
    \text{Mass at}\;&P=P_{\max}\rightarrow\quad &\mathcal{F}_P(P)&=\delta(P-P_{\max})\rightarrow\quad&\mathcal{S}=0
\end{align*}
As we are interested in the minimum of $\mathcal{S}$, the probability density $\mathcal{F}_P(P)=\delta(P-P_{\max}$) is of our primary interest. This probability density function is physically interpreted as the power collected being constant with value $P=P_{\max}$ for any pointing error. Hence, the continuous PDF has been converted to a discrete probability function. Furthermore, the minimality condition of $\mathcal{S}=0$ is also met by all the PDFs satisfying
\begin{equation}\label{eq_FP}
    \mathcal{F}_P(P)=\delta(P-P')\quad \text{for} \quad P'\geq P_{\rm th}
\end{equation}
Now the problem remains of finding a far-field irradiance that, in combination with the pointing jitter probability density function (e.g. Rayleigh distribution in Eq.~\eqref{eq_rayleigh}), gives the probability function in Eq.~\eqref{eq_FP} for the collected power. Finally, it can be seen directly from the definition of $\mathcal{S}$ in Eq.~\eqref{eq_S} that any function
\begin{equation}\label{eq_FPminmath}
    \mathcal{F}_P(P)=\begin{cases}
        0\quad &\text{if}\quad 0\leq P< P_{\text{th}}\\
        n(P) &\text{else}
    \end{cases}
\end{equation}
also satisfies the condition $\mathcal{S}=0$, where $n(P)$ is any positive function satisfying the normalization condition in Eq.~\eqref{eq_norm}
\begin{equation*}
    \int_{P_{\rm th}}^{P_{\max}} n(P) = 1
\end{equation*}
Now, for $\mathcal{F}_P(P)$ to be zero for $P\in[0, P_{\text{th}})$ as required by the solution in Eq.~\eqref{eq_FPminmath}, according to the definition of $\mathcal{F}_P(P)$ in Eq.~\eqref{eq_FPdef1}
\begin{equation*}
    \iint_{-\infty}^\infty f_\text{XY}(x_0,y_0)\;\delta[P-g(x_0,y_0)]\;dx_0\;dy_0=0\quad \text{for}\quad 0\leq P< P_\text{th}
\end{equation*}
and hence 
\begin{equation*}
    f_\text{XY}(x_0,y_0)=0\quad\quad  \forall (x_0,y_0)\in \mathbb{R}^2 
\end{equation*}
or
\begin{equation*}
    \delta[P-g(x_0,y_0)]=0\quad\quad 0\leq P\leq P_\text{th}, \;\;\forall (x_0,y_0)\in \mathbb{R}^2
\end{equation*}
The first condition above is not satisfied for a continuous unbounded pointing jitter (e.g. Rayleigh distribution in Eq.~\eqref{eq_rayleigh}). Regarding the second condition, as $g(x_0,y_0)>0$, this last condition means that the power collected is non-zero, $g(x_0,y_0)\geq P_{\text{th}}$, for any pointing jitter error $(x_0,y_0)\in \mathbb{R}^2$. Under this condition, the total power of the beam would not be bounded, and therefore this solution is not physically possible (see Proof 1 in \S\ref{app_proofs}).

Considering the above, we have now demonstrated that the $\mathcal{F}_P(P)$ given in Eq.~\eqref{eq_FPminmath} can not be fulfilled in general by a finite power far-field irradiance when considering an unbounded pointing jitter PDF. Therefore, now we are looking for a $\mathcal{F}_P(P)$ that gives an $\mathcal{S}>0$. Although the condition  $g(x_0,y_0)\geq P_{\text{th}}$ for $\forall (x_0,y_0)\in \mathbb{R}^2$ can not be met, the best solution would be one in which the condition $g(x_0,y_0)\geq P_{\text{th}}$ is met with the largest probability possible. In other words, we are looking for a function $g(x_0,y_0)$ that maximizes the probability of the collected power being above the given threshold $P_\text{th}$. Hence, the probability
\begin{equation}\label{eq_maximizePr}
    \Pr\bigl\{g(x_{0},y_{0})\ge P_{\rm th}\bigr\}
\;=\;
\iint_{\{(x_{0},y_{0}):\,g(x_{0},y_{0})\ge P_{\rm th}\}}
f_\text{XY}(x_{0},y_{0})\,dx_{0}\,dy_{0},
\end{equation}
needs to be maximized. At this point, it is important to prove that the finite power condition in Eq.~\eqref{eq_finitepower} implies the finiteness of the bidimensional integral of $g(x_0,y_0)$ through the relation in Eq.~\eqref{eq_gx0y0}. The reader is referred to Proof 2 in \S\ref{app_proofs} for the proof that given that the power of the beam is finite (see Eq.~\eqref{eq_finitepower}), then
$$
\iint_{\mathbb{R}^2}\;g(x_0,y_0)\;dx_0\,dy_0\leq G
$$
where $G$ is finite. As the PDF of the pointing jitter $f_\text{XY}(x_0,y_0)$ is considered radially decreasing (see Eq.~\eqref{eq_rayleigh}), we get that the maximization in Eq.~\eqref{eq_maximizePr} is achieved when the area described by the condition ${\{(x_{0},y_{0}):\,g(x_{0},y_{0})\ge P_{\rm th}\}}$ is a circular region in two-dimensional space with the largest possible radius. In Proof 3 of \S~\ref{app_proofs} we show that the maximization of this area is obtained by a circular step function of height $P_\text{th}$,
\begin{equation}\label{eq_gopt}
    g(x,y)=
    \begin{cases}
    P_{\rm th},&x^2+y^2\le r_0^2,\\
    0,&\text{otherwise},
    \end{cases}
\end{equation}

\begin{figure}
    \centering
    \includegraphics[width=0.65\linewidth]{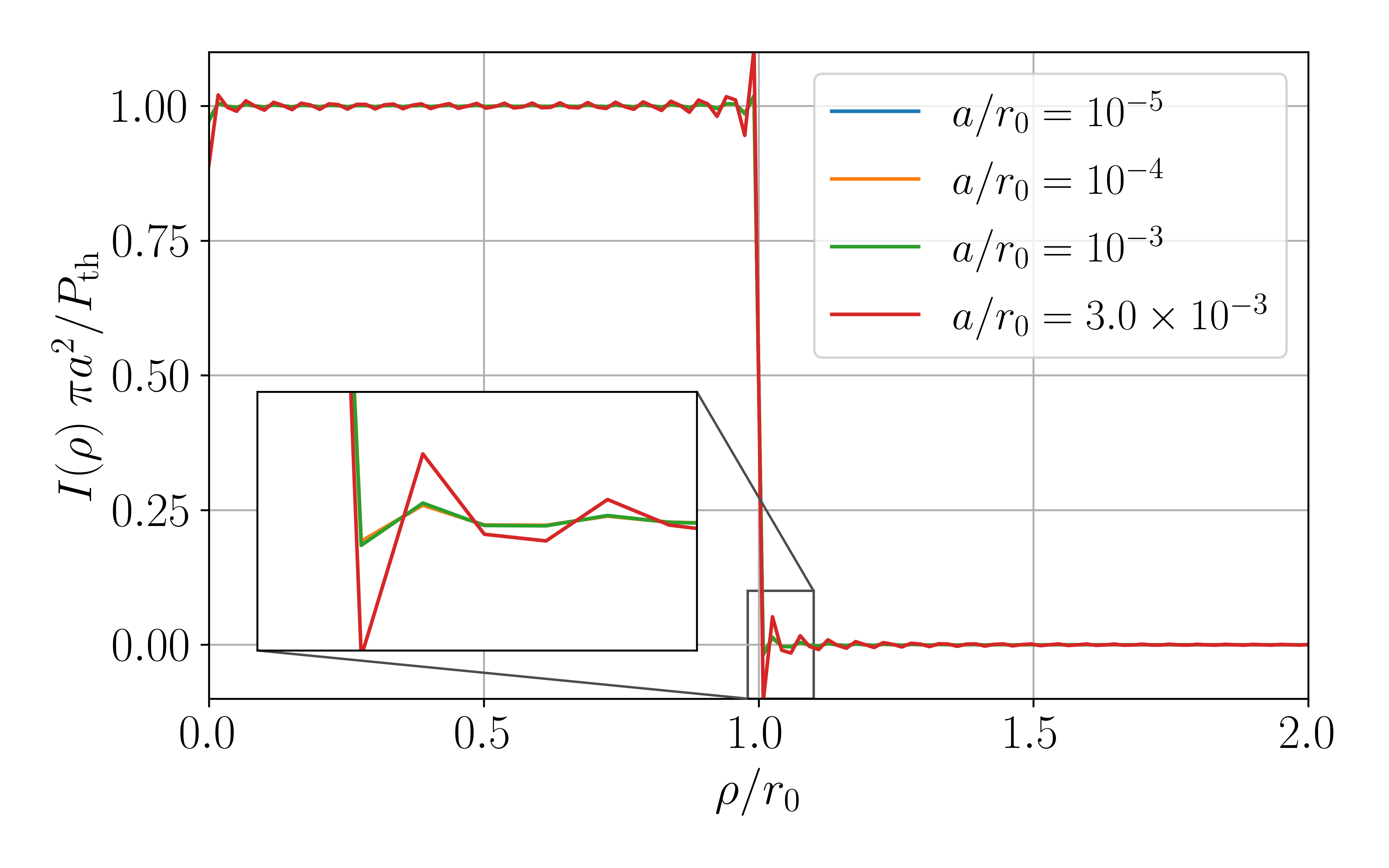}
    \caption{Optimum normalized far-field irradiance obtained through the numerical inverse Hankel transform of $\tilde I$ obtained through Eqs.~\eqref{eq_Ik2}-\eqref{eq_hankel}.}
    \label{fig_Hankeltransform}
\end{figure}

Having established the condition that $g(x_0,y_0)$ must satisfy in order to minimize $\mathcal{S}$, we can now determine the corresponding optimum far-field irradiance $I(x,y)$ that produces such a power collection profile. To this end, we use the relation between these functions given in Eq.~\eqref{eq_gx0y0}. By applying the Fourier-transform properties of Eq.~\eqref{eq_gx0y0}, we obtain, in momentum space,
\begin{equation}\label{eq_Ik}
     \tilde I(k_x,k_y)=\dfrac{\tilde g(k_x,k_y)}{\tilde{\mathcal A}(k_x,k_y)}
\end{equation}
where $\tilde \nonumber$ refers to the Fourier transform. Therefore, for a given aperture $\mathcal{A}$ and considering the optimum power collection function in Eq.~\eqref{eq_gopt}, we can obtain the optimum far field irradiance operating in the Fourier domain. Considering a circular aperture $\mathcal A$, the Fourier components of the irradiance field are (see Proof 4 in \S~\ref{app_proofs})
\begin{equation}\label{eq_Ik2}
    \tilde{I}(k)
    =\;P_{\rm th}\;\frac{r_0}{a}
    \;\frac{\mathcal J_1(|k|\,r_0)}{\mathcal J_1(|k|\,a)}.
\end{equation}
where $a$ is the radius of the aperture and $\mathcal J_1$ is the Bessel function of the first kind. The inverse Fourier transform (which in polar coordinates results in the inverse Hankel transform of the first kind due to the azimuthal symmetry of the problem) of this ratio is a function in polar coordinates given by
\begin{equation}\label{eq_hankel}
    I(\rho) = \int_{\mathbb{R}^+}\tilde{I}(k)\;J_1(k\rho) \;\rho\;d\rho
\end{equation}
However, the general solution can not be found in a closed form due to the divergences created by the zeros of the Bessel function in the denominator of Eq.~\eqref{eq_Ik2}. For the small circular aperture approximation, the inverse Hankel transform involved can be computed (see Proof 4 in \S~\ref{app_proofs}), yielding the far-field irradiance
\begin{equation*}
    I(\rho)=\begin{cases}
        \dfrac{2P_\text{th}r_0}{a^2} &\rho\leq r_0\\
        0 &\text{otherwise}
    \end{cases}
\end{equation*}
which is the flat-top beam (up to a power normalization constant). Imposing the available finite power condition of Eq.~\eqref{eq_finitepower} we get that the flat-top beam is
\begin{equation*}
    I(\rho)=\begin{cases}
        \dfrac{P_\text{th}}{\pi a^2} &\rho\leq r_0, \quad \text{where}\quad r_0 = a\sqrt {\dfrac{P_0}{P_\text{th}}}\\
        0 &\text{otherwise}
    \end{cases}
\end{equation*}
For larger apertures, Eq.~\eqref{eq_Ik} can be numerically computed, although as shown in Fig.~\ref{fig_Hankeltransform}, instabilities arise as the ratio between the aperture radius $a$ and the beam size $r_0$ increases. It can be seen in the numerical results that the inverse Hankel transform oscillates strongly, giving unphysical negative irradiance values when the ratio $a/r_0$ is increased. However, it is shown that the shapes obtained numerically for bigger apertures closely resemble the flat-top beam obtained in the small aperture limit. For big $a/r_0$ ratios, the oscillations obtained mathematically can be understood considering that the variations of the irradiance across the aperture can still compensate each other, yielding the same total power collected by the aperture, i.e. $P_\mathrm{th}$.

\section{Analytical solutions of some beam shapes}\label{sec_smallap}
Building upon the previous section’s finding that a flat-top profile is the optimum beam shape for mitigating pointing jitter, this section derives the analytical performance results, denoted by $\mathcal S$, for several key beam shapes. The Gaussian beam is examined first, because of its widespread application in free-space optical communications \cite{toyoshima_optimum_2002, badas_seidel_2025, yura_optimum_1995, farid_outage_2007}. Subsequently, the study extends to super-Gaussian beam shapes, as they represent a versatile family of functions that converge toward the flat-top profile in the limit of infinite order. Finally, the flat-top beam itself is analyzed in detail, as it represents the ideal solution under the conditions previously identified.

Due to the complexity of the mathematical operations involved, this analysis adopts the small-aperture hypothesis and assumes a Rayleigh density function for the pointing jitter, as expressed in Eq.~\eqref{eq_rayleigh}. Under the assumption that the receiver telescope aperture is sufficiently small, the irradiance variation across the aperture becomes negligible. Consequently, the power collected as a function of the radial distance $\rho_0$ between the receiver and the beam center can be expressed as in Eq.~\eqref{eq_grho0}.

\subsection{Gaussian beam}\label{subsec_ch11_gaussian}
We begin our derivation by analyzing the Gaussian beam profile, which serves as the benchmark for free-space optical communications systems due to its role as the fundamental mode in laser resonators and single-mode fibers usually employed in optical terminals. At the receiver aperture plane, a Gaussian beam with a total power $P_0$ is given by
\begin{equation}\label{eq_gaussian}
    I(\rho)=\frac{2P_{0}}{\pi w^2}\exp \left({\frac {-2\rho^{2}}{w^{2}}}\right).
\end{equation}
Under the small-aperture approximation established previosuly in Eq.~\eqref{eq_grho0}, the power collected by a receiver of area $A$ is proportional to the local irradiance $P(\rho)=AI(\rho)$. To determine the statistics of the received power under the influence of pointing jitter, we first find the inverse relationship expressing the radial displacement as a function of received power
\begin{equation*}
    \rho(P) 
    = \sqrt{\dfrac{w^2}{2}\ln\left( \dfrac{2AP_0}{P\pi w^2}\right)} 
\end{equation*} 
Assuming a Rayleigh distribution for the pointing jitter (as defined in Eq.~\eqref{eq_rayleigh}), the probability density function of the received power, $\mathcal{F}_\text{P,Gaussian}(P)$, is computed via the transformation of random variables in Eq.~\eqref{eq_FPdef2}
\begin{equation*}\begin{split}
    \mathcal{F}_\text{P,Gaussian}(P) =& \left|{\frac {d}{dP}}\rho(P)\right|\cdot f_\mathrm{R}{\big (}\rho(P){\big )} \\
    \propto& \dfrac{1}{P\sigma^2 } w^2 2^{-\beta-3} \pi ^{\beta-1} \left(\frac{A P_0}{P w^2}\right)^{-\beta} \quad\text{with}\quad \beta=\dfrac{w^2}{4\sigma^2}
\end{split}\end{equation*}
By imposing the normalization condition $\int_0^{P_{\max}} \mathcal{F}_{\text{P,Gaussian}}(P)\;dP=1$, where $P_{\max}=2AP_{0} / \pi w^2$ corresponds to the peak of the irradiance, the complete PDF is found to be 
\begin{equation*}\begin{split}
    \mathcal{F}_\text{P,Gaussian}(P) =& \dfrac{1}{P\sigma^2 } w^2 2^{-\beta-2} \pi ^{\beta} \left(\frac{A P_0}{P w^2}\right)^{-\beta} \quad\text{with}\quad \beta=\dfrac{w^2}{4\sigma^2}
\end{split}\end{equation*}
Since the beam profile is constrained to a Gaussian shape, we treat this as a parametric optimization problem where the beamwidth $w$ is the decision variable\footnote{The average power captured can be obtained from the PDF above, and is given by $\overline{P}=2AP_0/(\pi w^2+4\pi\sigma^2)$ which has a maximum $\max\overline{P}=AP_0/(2\pi\sigma^2)$ in $w\to0$. In the regime $w\to0$ the approximation of the beam width being bigger than the receiver aperture is no longer valid. In this regime, the average captured power is given by $\overline{P}=P_0\,[1-\exp(-A/2\pi\sigma^2)]$, that under the approximation that the aperture is smaller than the pointing jitter, i.e. $A\ll\sigma^2$, yields the maximum above when expanding $\overline P$ in power series.}. The objective is to minimize the performance metric $\mathcal S$, defined as the probability that the received power falls below a given threshold $P_\mathrm{th}$
\begin{equation*}
\begin{split}
    \mathcal{S} (w) =& \int_0^{P_\text{th}} \mathcal{F}_\text{P,Gaussian}(P) \;dP=\left(\dfrac{2}{\pi}\dfrac{A P_0}{w^2 P_\text{th}}\right)^{-\beta} 
\end{split}    
\end{equation*}
To find the optimum beamwidth, we evaluate ${d\mathcal{S}}/{dw}=0$, that has the only non-trivial root ($w\neq0)$
\begin{empheq}[box=\fbox]{equation}\label{eq_woptGaussian}
    w_\text{opt,Gaussian}=\sqrt{\dfrac{{2} A {P_0}}{{e \pi }{P_\text{th}}}}
\end{empheq}
Under this condition the minimum $\mathcal{S}$ will be given by
\begin{empheq}[box=\fbox]{equation}\label{eq_SminGaussian}
    \min{\mathcal{S}_{\text{Gaussian}}}=\exp\left(-\dfrac{A\,P_0}{2\,e\,\pi \,P_\text{th}\,\sigma^2}\right)
\end{empheq}
The relationships between the optimum beamwidth and the performance metric $\mathcal S$ as a function of the system parameters $\{A,P_0,P_\text{th},\sigma\}$ are further illustrated in Fig.~\ref{fig_G_SG_FT}.

\subsection{Super-Gaussian beams}
While the Gaussian beam is the most used in free-space optical communications, it is not necessarily the most robust profile against the detrimental effects of pointing jitter. As demonstrated by the variational analysis in \S~\ref{sec_variational_FP}, flat-top-like distributions offer superior performance by providing a more uniform irradiance over a wider area. Since an ideal flat-top profile is not strictly attainable in practice, due to its discontinuity, it is natural to consider finite-order super-Gaussian beams as realistic approximations. These functions are characterized by a steepness parameter, or order $n_\mathrm{SG}$, which allows them to continuously approximate the stepwise flat-top function as $n_\mathrm{SG}$ increases—a property that has led to their use in several applications, e.g., inertial confinement fusion \cite{deng_pure-phase_1996, gillen-christandl_comparison_2016, dickey_laser_2018}.

The radially varying irradiance profile of a super-Gaussian beam is mathematically defined as
\begin{equation*}
    I(\rho) = \dfrac{P_0 \,4^{1/n_\mathrm{SG}}}{\pi\, w^2 \,\Gamma\left[(2+n_\mathrm{SG})/n_\mathrm{SG}\right]}\;\exp \left[ { - 2{{\left( {\frac{\rho}{w}} \right)}^{n_\mathrm{SG}}}} \right],
\end{equation*}
where $n_\mathrm{SG}$ is the beam order, $P_0$ is the total power, and $\Gamma (x)$ is the Gamma function. Notably, for $n_\mathrm{SG}=2$, this expression reduces to the standard Gaussian irradiance in Eq.~\eqref{eq_gaussian}. Under the small-aperture approximation, the power collected at a radial offset $\rho$ is given by $P(\rho)=A\;I(\rho)$.

Since the received power $P(\rho)$ is a strictly monotonically decreasing function of the radial displacement ($dP/d\rho\leq 0$) for all $\rho\in[0,\infty)$, there exists a unique, one-to-one mapping between the power and the pointing error. Consequently, the event where the received power falls below a critical threshold $P_\text{th}$—defined as the performance parameter $\mathcal S$—is statistically equivalent to the event where the radial jitter exceeds a corresponding spatial displacement $\rho_\text{th}$. By solving $P(\rho_\text{th})=P_\text{th}$, we define the boundary of the outage probability zone in the spatial domain. Because larger pointing errors result in lower power, the probability of an outage is captured by the tail of the jitter distribution. Thus, $\mathcal S$ can be calculated directly using the complementary cumulative distribution function of the Rayleigh pointing jitter, expressed as
\begin{equation*}
    \mathcal{S}=\int_{\rho_{\text{th}}}^\infty  f_\mathrm{R}(\rho)  \;d\rho=\exp\left(-\dfrac{\rho_{\text{th}}^2}{2\sigma^2}\right).
\end{equation*}
where $\rho_{\text{th}}$ is the point at which the power collected $P(\rho)$ meets the power threshold $P(\rho_{\text{th}})=P_{\text{th}}$. To obtain this value,
\begin{equation*}\begin{split}
    P_{\text{th}}= \dfrac{A P_0 4^{1/n_\mathrm{SG}}}{\pi w^2 \Gamma\left[(2+n_\mathrm{SG})/n_\mathrm{SG}\right]}\;&\exp \left[ { - 2{{\left( {\frac{\rho_{\text{th}}}{w}} \right)}^{n_\mathrm{SG}}}} \right]\\
    &\xrightarrow[]{}\rho_{\text{th}}=\dfrac{w}{2^{1/n_\mathrm{SG}}}  \left[-\ln \left(\dfrac{4^{-1/n_\mathrm{SG}}\pi P_{\text{th}}w^2\Gamma\left[(2+n_\mathrm{SG})/n_\mathrm{SG}\right]}{A\;P_0}\right) \right]^{1/n_\mathrm{SG}}
\end{split}\end{equation*}
This transformation simplifies the analysis by shifting the integration from the complex power domain to the well-defined statistical properties of the pointing error. To achieve the best system performance, we seek to minimize the outage probability $\mathcal S$ by maximizing the radial tolerance $\rho_\text{th}$. Physically, this represents finding the beamwidth $w$ that allows for the largest possible pointing error before the received power drops below $P_\text{th}$. By applying the first-order optimality condition $dw/d\rho_\text{th}=0$, we identify the optimum beamwidth as
\begin{empheq}[box=\fbox]{equation} \label{eq_woptSG}
    w_{\text{opt,SG}}=\sqrt{\dfrac{AP_0}{\pi P_\text{th}\Gamma\left[(2+n_\mathrm{SG})/n_\mathrm{SG}\right]}}\left(\dfrac{2}{e}\right)^{1/n_\mathrm{SG}}.
\end{empheq}
In the standard Gaussian case where $n_\mathrm{SG}=2$, this expression reduces to Eq.~\eqref{eq_woptGaussian}. At this optimum beamwidth, the maximum allowable radial displacement is given by
\begin{equation*}
    \rho_\text{th}=\sqrt{\dfrac{AP_0}{\pi P_\text{th}\Gamma\left[(2+n_\mathrm{SG})/n_\mathrm{SG}\right]}} \left(\dfrac{2}{n_\mathrm{SG} e}\right)^{1/n_\mathrm{SG}}.
\end{equation*}
Consequently, the minimized performance parameter for the super-Gaussian beam is determined by substituting this maximized $\rho_\text{th}$ into the Rayleigh complementary cumulative density function, yielding
\begin{empheq}[box=\fbox]{equation}\label{eq_sminSG}
    \min{\mathcal{S}_\text{SG}}=
       \exp\left(-\dfrac{\gamma_{\text{SG}}^2}{2}\right)
\end{empheq}
with 
\[
\gamma_{\text{SG}}=\sqrt{\dfrac{AP_0}{\pi\sigma^2 P_\text{th}\Gamma\left[(2+n_\mathrm{SG})/n_\mathrm{SG}\right]}} \left(\dfrac{2}{n_\mathrm{SG} e}\right)^{1/n_\mathrm{SG}}.
\]
Consistent with our previous derivations, this result matches the solution for the standard Gaussian beam in Eq.~\eqref{eq_SminGaussian} when $n_\mathrm{SG}=2$. A comparison between the optimum super-Gaussian beams of several orders and the standard Gaussian beam is illustrated in Fig.~\ref{fig_G_SG_FT}.

\subsection{Flat-top beam}

\begin{figure}
    \centering
    \includegraphics[width=\linewidth]{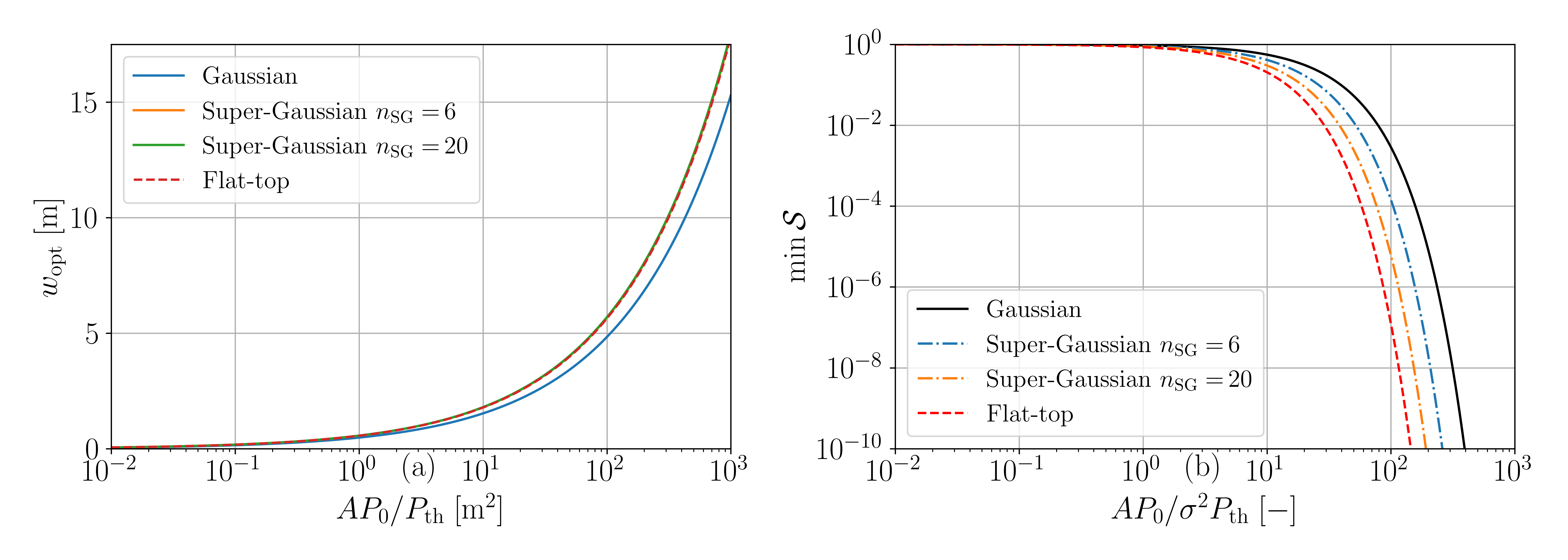}
    \caption{Comparison of optimum Gaussian, super-Gaussian, and flat-top beams. The optimum beamwidth expressions and the corresponding performance metric $\mathcal S$ are shown as functions of the system parameters $\{A,P_0,P_\text{th},\sigma\}$.}
    \label{fig_G_SG_FT}
\end{figure}

\begin{figure}
    \centering
    \includegraphics[width=0.5\linewidth]{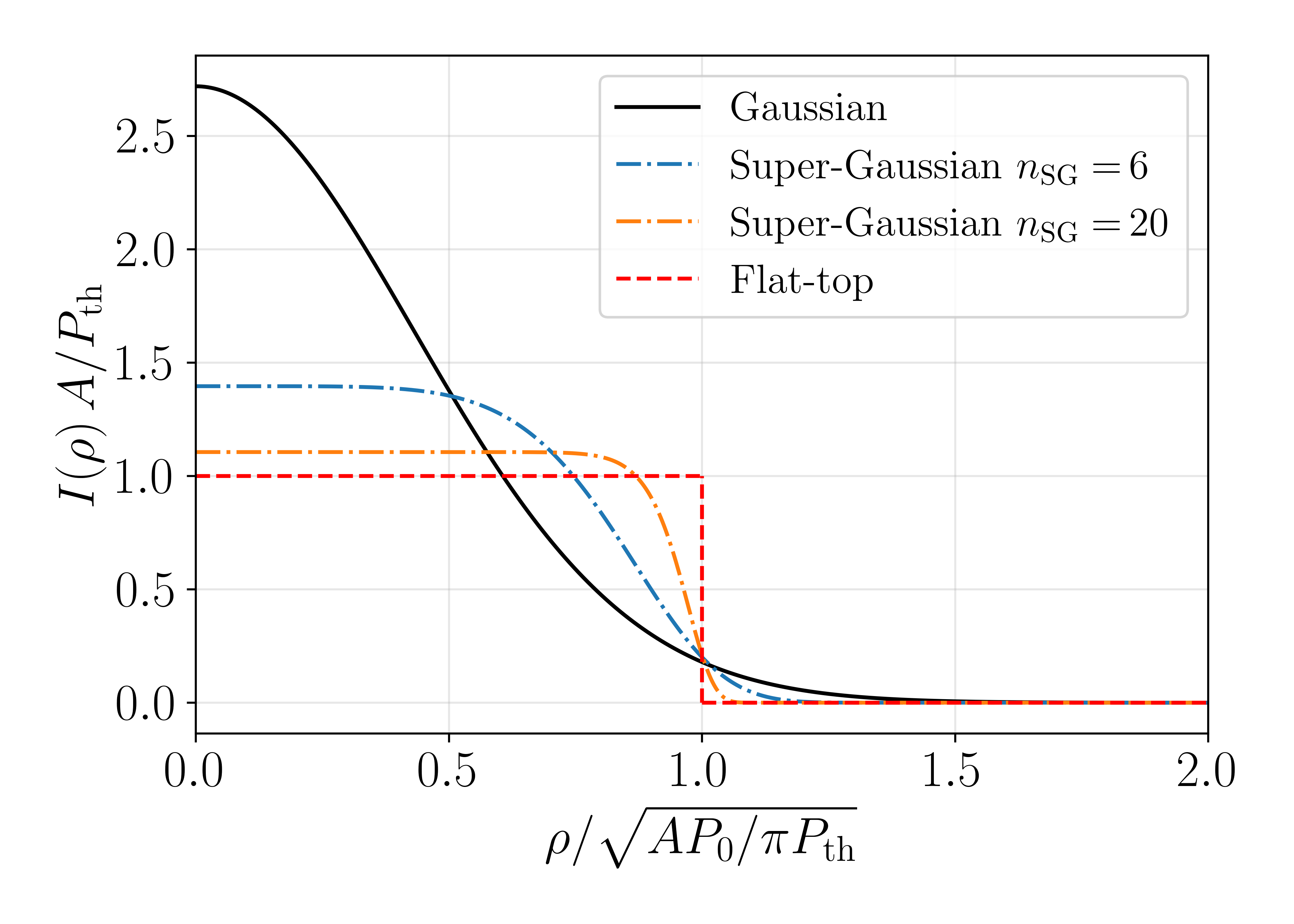}
    \caption{Optimum far-field irradiances (normalized) for different Gaussian, super-Gaussian, and flat-top beams.}
    \label{fig_fields}
\end{figure}

In the previous section, we introduced super-Gaussian beams and showed that, as the order increases, their transverse intensity profiles approach an ideal flat-top distribution. Moreover, in \S~\ref{sec_variational_FP} we demonstrated—under suitable approximations—that the flat-top profile constitutes the optimum solution for mitigating the effect of the pointing jitter. Hence, the remaining question is to determine which flat-top solution is optimum. To address this, we consider a flat-top beam characterized by a total power $P_0$ and a width $w$. These parameters fully specify the profile and allow us to investigate how the performance depends on the choice of beam width under fixed total power. Mathematically, the flat-top beam is described by the intensity profile
\begin{equation*}
    I(\rho) = \begin{cases}
    \dfrac{P_0}{\pi w^2}\quad &\rho\leq w \\
    0 \quad &\text{else}
    \end{cases}
\end{equation*}
As in the Gaussian-beam case, determining the optimum beam width $w$ constitutes an optimization problem. 
For a flat-top beam, the received power can take only two possible values depending on whether the receiver aperture lies inside or outside the illuminated region. Consequently, the probability density function of the received power becomes discrete, with values $\{AP_0/(\pi w^2), 0\}$ occurring with probabilities $\{1-\mathcal S,\mathcal S\}$, respectively\footnote{The average power captured can be obtained from the discrete probability distribution above, and is given by $\overline{P}=AP_0(1-e^{-\frac{w^2}{2 \sigma^2}})/(\pi w^2)$ which has a maximum $\max\overline{P}=AP_0/(2\pi\sigma^2)$ in $w\to0$, as obtained for the Gaussian beam in this limit.}. Here
\begin{equation*}
    \mathcal S = \int_w^\infty f_\mathrm{R}(\rho)\, d\rho=\exp\left(-\dfrac{w^2}{2\sigma^2}\right)
\end{equation*}
represents the probability that the radial pointing error exceeds the beam radius $w$. For the Rayleigh distributed pointing jitter considered, this probability corresponds to the complementary cumulative density function evaluated at $w$. In principle, increasing the beam width $w$ reduces $\mathcal S$, thereby decreasing the probability that the receiver falls outside the beam and improving system performance. However, increasing $w$ simultaneously reduces the received power level $AP_0/(\pi w^2)$. If this power falls below the threshold $P_\text{th}$, the received power is always insufficient, and therefore the probability of collecting power below the threshold becomes unity, i.e. $\mathcal S=1$. This situation occurs when
\begin{equation*}
    \dfrac{AP_0}{\pi w^2}<P_\text{th} \rightarrow w>\sqrt{\dfrac{AP_0}{ \pi P_\text{th}}}
\end{equation*}
Therefore, the optimum performance—corresponding to the minimum cumulative probability of the received power falling below the threshold $P_\text{th}$—is achieved when the beam width is chosen such that the received power inside the beam is exactly equal to the threshold value. Under this condition, the optimum intensity profile becomes
\begin{equation*}
    I(\rho) = \begin{cases}
    \dfrac{P_\text{th}}{A}\quad &\rho<w_{\text{opt,FT}} \\
    0 \quad &\text{else}
    \end{cases}
\end{equation*}
where the optimum beam width is given by
\begin{empheq}[box=\fbox]{equation} \label{eq_woptFT}
    w_{\text{opt,FT}}=\sqrt{\dfrac{AP_0}{ \pi P_\text{th}}}
\end{empheq}
The corresponding minimum probability that the received power falls below $P_\text{th}$, which serves as the performance metric, is
\begin{empheq}[box=\fbox]{equation}\label{eq_sminFT}
    \min{\mathcal{S}_\text{FT}}=
       \exp\left(-\dfrac{\gamma_{\text{FT}}^2}{2}\right)
\end{empheq}
where
\[\gamma_{\text{FT}}=\sqrt{\dfrac{AP_0}{ \pi \sigma^2P_\text{th}}}\]
This result establishes a direct relation between the maximum achievable performance—in terms of outage probability—and the system parameters: threshold power $P_\text{th}$, receiver aperture area $A$, pointing jitter $\sigma$, and total transmitted power $P_0$. In Fig.~\ref{fig_G_SG_FT}, the optimum beamwidth and the corresponding performance metric $\mathcal{S}$, as functions of the parameters $\{A,P_0,P_\text{th},\sigma\}$, are shown together with the optimum solutions obtained for Gaussian and super-Gaussian beams. Furthermore, Fig.~\ref{fig_fields} shows the far-field irradiances for the different beams analyzed. It has also been checked that the ring flat-top (a donut-like top hat with a hole in the center) profile results in a worse performance compared to the standard flat-top function (see \S~\ref{app_ring}).

In the presence of an \textit{unknown} boresight or a static pointing error, the pointing statistics are described by a Rice distribution (if the static error is known, then the system should be recalibrated, and the effect of the remaining pointing jitter could be mitigated by means of the beams proposed in this work). Nevertheless, the resulting optimization problem remains unchanged: the objective is to maximize the central region of the far field where the collected power exceeds the specified power threshold. Since the area covered by the flat-top beam includes that of a conventional Gaussian beam, the minimum achievable $\mathcal{S}$ will remain smaller for the flat-top, but will differ from the value obtained under the Rayleigh distribution assumption discussed above. By similar reasoning, if the estimation of the pointing jitter $\sigma^2$ is inaccurate, the optimization still yields the same result, since the optimum beamwidths depend only on the ratio $AP_0/P_\mathrm{th}$ and not on the pointing jitter itself. If the actual pointing jitter is smaller than expected, the achieved minimum $\mathcal{S}$ will also be smaller (see Fig.~\ref{fig_G_SG_FT}).

Another practical aspect of interest arises when an overdimensioning coefficient $\mathcal{C} > 1$ is applied to the value of $P_\mathrm{th}$ in link performance estimations. Equivalently, this can be interpreted as applying the same scaling factor to $\sigma^2$, or, conversely, as introducing an underdimensioning factor $\mathcal{C}^{-1}$ to the parameters $\{A, P_0\}$. In this case, the computed far-field beam widths for Gaussian, super-Gaussian, and flat-top beams are all scaled by a factor $1/\sqrt{\mathcal{C}}$ (see Eqs.~\eqref{eq_woptGaussian}, \eqref{eq_woptSG}, \eqref{eq_woptFT}). Consequently, the resulting value of $\mathcal{S}$ is no longer the minimum achievable for a given beam shape. Specifically, for a Gaussian beam, the achieved $\mathcal{S}$ becomes $(\min \mathcal{S}_\mathrm{Gaussian})^\Sigma$, where $\Sigma = (1 + \ln \mathcal{C}) / \mathcal{C}$. For a super-Gaussian beam, it becomes $(\min \mathcal{S}_\mathrm{SG})^\Omega$, where 
\[
\Omega = \dfrac{1}{\mathcal C} \left[1 + \frac{n_\mathrm{SG} \ln \mathcal{C}}{2} \right]^{2/n_\mathrm{SG}}.
\]
For a flat-top beam, the corresponding expression is $(\min \mathcal{S}_\mathrm{FT})^{1/\mathcal{C}}$. In all cases, the system performance degrades relative to the respective optima due to the overestimation of the power threshold. It is therefore of interest to determine the value of the overdimensioning coefficient for which the performance of the Gaussian and flat-top beams becomes identical. This can be obtained by solving for $\mathcal{C}$ such that
\(
(\min \mathcal{S}_\mathrm{Gaussian})^\Sigma = (\min \mathcal{S}_\mathrm{FT})^{1/\mathcal{C}}.
\)
Numerical evaluation shows that this occurs at $\mathcal{C} \approx 5.57$. Thus, when such an overdimensioning factor is applied, the resulting performance $\mathcal{S}$ of both Gaussian and flat-top beams is the same, and the flat-top beam offers no performance advantage.

\begin{figure}
    \centering
    \includegraphics[width=0.5\linewidth]{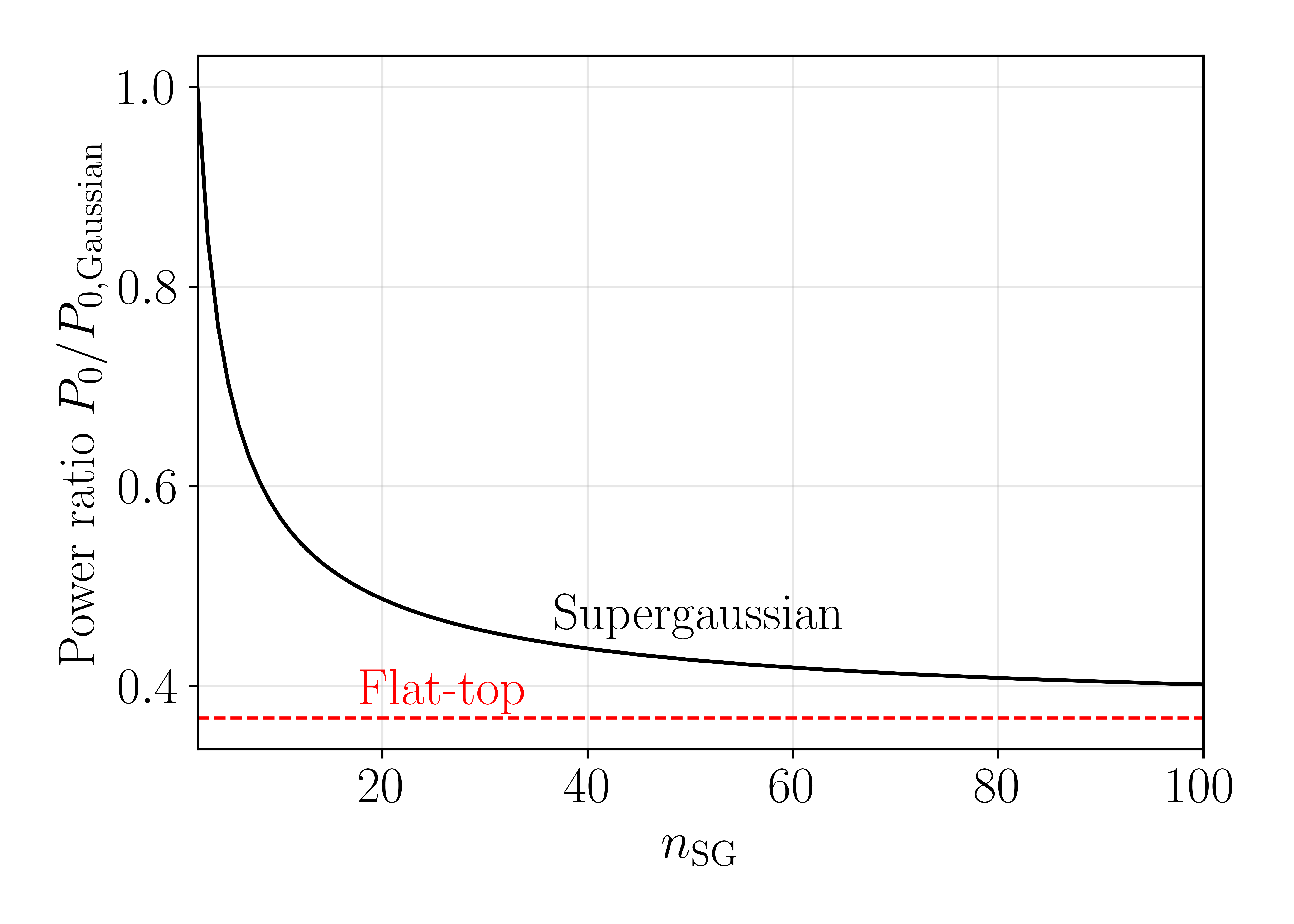}
    \caption{Power savings obtained with super-Gaussian and flat-top beams. Ratio $P_0/P_{0,\mathrm{Gaussian}}$ of the total power required to achieve the same performance $\mathcal S$ as a function of the super-Gaussian order $n_\mathrm{SG}$. The horizontal dashed line indicates the limiting case corresponding to the optimum flat-top beam.}
    \label{fig_gain_n_SG_FT}
\end{figure}

To quantify the power savings provided by super-Gaussian and flat-top beams relative to the fundamental Gaussian beam, Fig.~\ref{fig_gain_n_SG_FT} shows the reduction in required transmitted power for a fixed performance level $\mathcal{S}$. The vertical axis represents the ratio $P_0/P_{0,\mathrm{Gaussian}}$, i.e., the total power required by a super-Gaussian beam to achieve the same performance as an optimally chosen Gaussian beam. Notably, this ratio is independent of the particular value of $\mathcal{S}$ selected and therefore represents an intrinsic measure of the relative efficiency of the different beam profiles. By mathematically equating the $\min\mathcal S$ of the super-Gaussian beams in Eq.~\eqref{eq_sminSG} with the Gaussian in Eq.~\eqref{eq_SminGaussian} the ratio \[\dfrac{P_0}{P_{0,\mathrm{Gaussian}}} = \dfrac{1}{e}\,\Gamma\left(\dfrac{2+n_\mathrm{SG}}{n_\mathrm{SG}}\right)\,\left(\dfrac{n_\mathrm{SG}\,e}{2}\right)^{2/n_\mathrm{SG}}.\] is obtained. As the super-Gaussian order $n_\mathrm{SG}$ increases, the required power decreases monotonically. This behavior reflects the increasing spatial uniformity of the beam profile and its progressive convergence toward the optimum flat-top distribution. In the limit of large $n_\mathrm{SG}$, the required power approaches that of the ideal flat-top beam, indicated by the horizontal dashed line, which represents the minimum power required to achieve the specified performance level. The power ratio required for the flat-top can be analytically obtained in a similar way to the super-Gaussian case, yielding a minimum $e^{-1}\approx36,8\%$ power ratio.

\section{Incoherent and coherent beam shaping}\label{sec_beamshaping}
In the previous sections, the optimum far-field irradiances have been analyzed, and their performance has been evaluated. However, a way of achieving these irradiances has not yet been proposed by optical means. To do so, in this section, several methods for approximating super-Gaussian and flat-top beams are discussed, and their performance is evaluated. 

Satellite optical communications are not the only application in which uniform irradiance of the beam is desirable. Other applications include inertial confinement fusion \cite{deng_pure-phase_1996}, laser material processing \cite{buske_advanced_2022,hayasaki_automatic_2023}, molecule manipulation \cite{dong_super-gaussian_2005} and addressing atomic qubits \cite{gillen-christandl_comparison_2016}. Leveraging on the optical systems proposed for achieving uniform irradiance fields, in this section we analyze two main physical means of approximating higher-order super-Gaussian beams. Firstly, incoherent light sources are considered, in which by means of dispersion, different transversal modes are superposed in irradiance to get an approximate flat-top beam. Secondly, a perfectly coherent light beam is assumed. In this case, a phase screen is designed by means of an iterative process to convert the original beam into an approximate flat-top beam. Although other approaches combining phase masks with spectral dispersion have also been proposed for obtaining flat-top beams  \cite{jiang_uniform_2011, deng_fresnel_2013}, these will not be investigated in this work.

\subsection{Incoherent light: Spatial mode superposition}\label{sec_incoherent}
Super-Gaussian and flat-top beams can be approximated in the far-field by superposing different free-space optical modes in irradiance. To superpose irradiance fields, however, it is necessary that the different modes are incoherently combined. Physically, this can be done by using different polarizations or wavelengths of light for each of the modes. Previous works have proposed the use of the superposition of orthogonally polarized modes to shape the far-field irradiance in a convenient way \cite{badas_optimum_2024}. This technique can be experimentally implemented by using simple polarization optics and diffractive optical elements \cite{badas_annular_2026}. However, as the polarization of light has only two degrees of freedom, only two modes can be superposed, strongly limiting the achievable variety of beam shapes with this technique. To overcome this limitation, different wavelengths of light can be used for each of the modes \cite{jacquard_enhancing_2025}. As the frequency of light is a continuous spectrum, there is no theoretical limit on the number of modes that we could combine using this technique. Experimentally, wavelength-based incoherent beam shaping can be achieved by means of dispersive elements and multi-plane light-converters, as has been demonstrated in Ref.~\cite{jacquard_enhancing_2025} for atmospheric turbulence mitigation. 

When using orthogonal polarization states to perform incoherent beam shaping through superposition, then polarization can no longer be used to modulate, multiplex, or separate transmit and receive optical paths on the optical communication terminal \cite{badas_opto-thermo-mechanical_2023}. Hence, in this case, wavelength would be used to separate these channels, as spatial mode is also being used for shaping the beam incoherently. When mode superposition is done by combining different wavelength beams with different modes, then polarization can be used for other purposes on the optical communication terminal. Not only that, in this case, other wavelength ranges away from the ones used for incoherent beam shaping can be used for modulation, multiplexing, and separation of transmit and receive optical paths.

When incoherently combining several spatial modes, the resulting irradiance is given by the superposition
\begin{equation}\label{eq_incoh}
    I(x,y)=\sum_{p=0}^{p_{\max}} \sum_{\ell=0}^{\ell_{\max}}c_{p\ell} I_{p\ell}(x,y)
\end{equation}
where $c_{p\ell}$ are the weights of the spatial modes $I_{p\ell}(x,y)$ considered, and the sum is done over the truncated modes, and $\{p,\ell\}$ are the indices of the spatial modes considered. These modes could be Laguerre-Gaussian (LG), Hermite-Gaussian (HG), or another set of spatial modes of our choice. The indices $\{p,\ell\}$ refer to radial and azimuthal indices of the LG modes, and the indices $\{n,m\}$ refer to the horizontal and vertical indices of the HG modes. As the number of modes considered increases, the field should get closer and closer to the desired beam shape. The coefficients $c_{p\ell}$ are calculated via a non-negative least-squares algorithm. More advanced ways of computing the optimum weights of the beams in incoherent superposition have also been proposed \cite{zhang_beam_2024}. In case polarization-based incoherent beam shaping is used, the summation in Eq.~\eqref{eq_incoh} can only be done over a maximum of two spatial modes \cite{badas_optimum_2024}.
\begin{figure}[!htb]
    \centering
    \includegraphics[width=\linewidth]{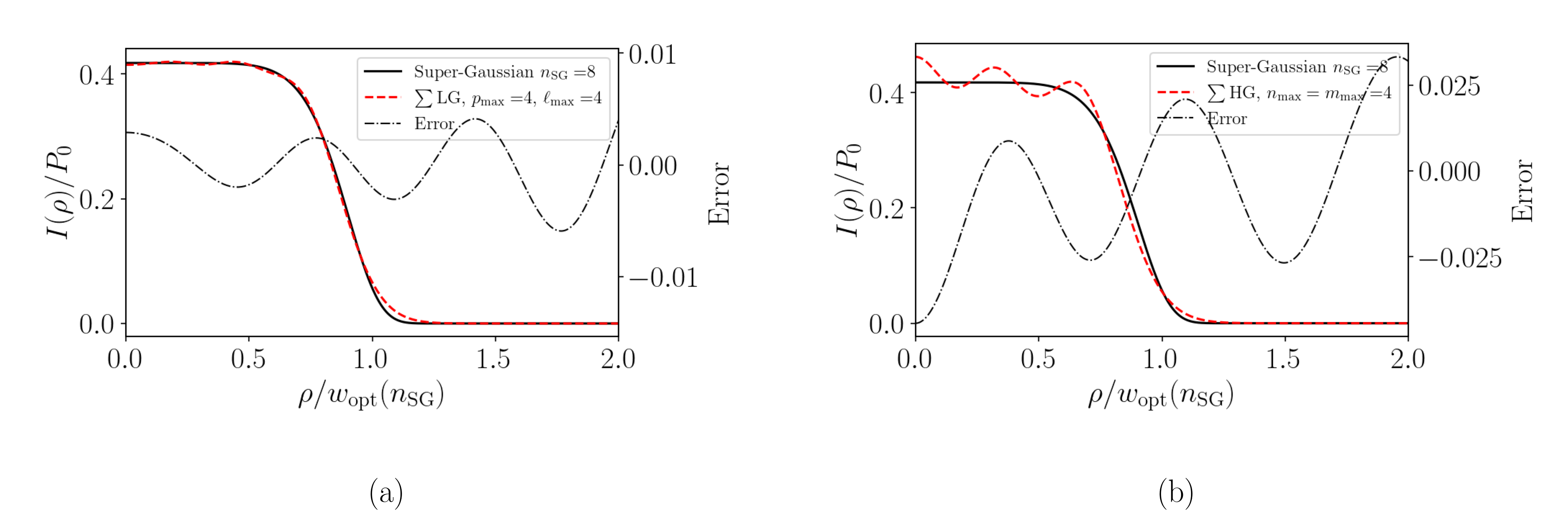}
    \caption{Incoherent beam shaping using (a) Laguerre-Gaussian and (b) Hermite-Gaussian modes, targeting the 8th order super-Gaussian beam.}
    \label{fig_incoh_fields}
\end{figure}

\begin{figure}[!htb]
    \centering
    \includegraphics[width=\linewidth]{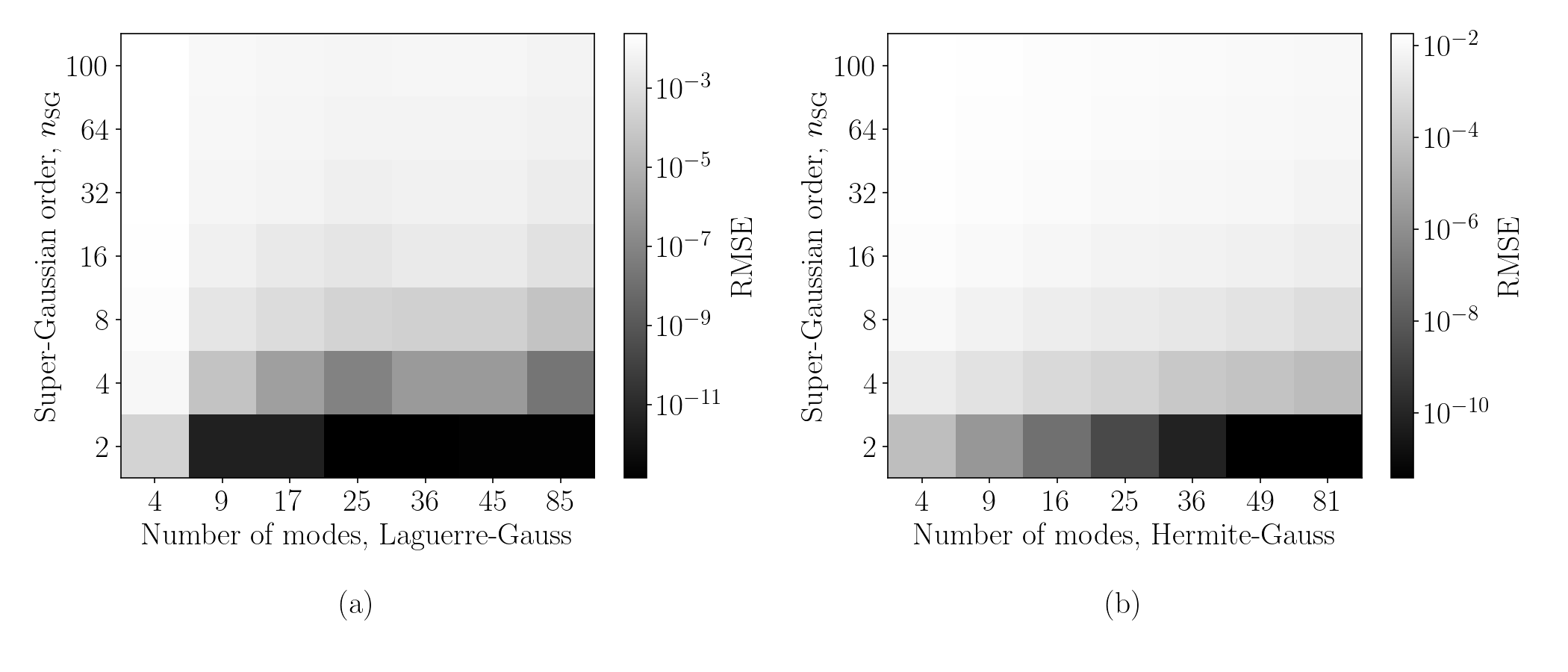}
    \caption{Root mean square error of the incoherent superposition of (a) Laguerre-Gaussian and (b) Hermite-Gaussian beams, when targeting several super-Gaussian order beams. The number of modes considered are $(p_{\max}+1)\times(\ell_{\max}+1)$ and $(n_{\max}+1)\times(m_{\max}+1)$ in the Laguerre-Gaussian and Hermite-Gaussian case, respectively.}
    \label{fig_incoh_rmse}
\end{figure}

\begin{figure}[!htb]
    \centering
    \includegraphics[width=\linewidth]{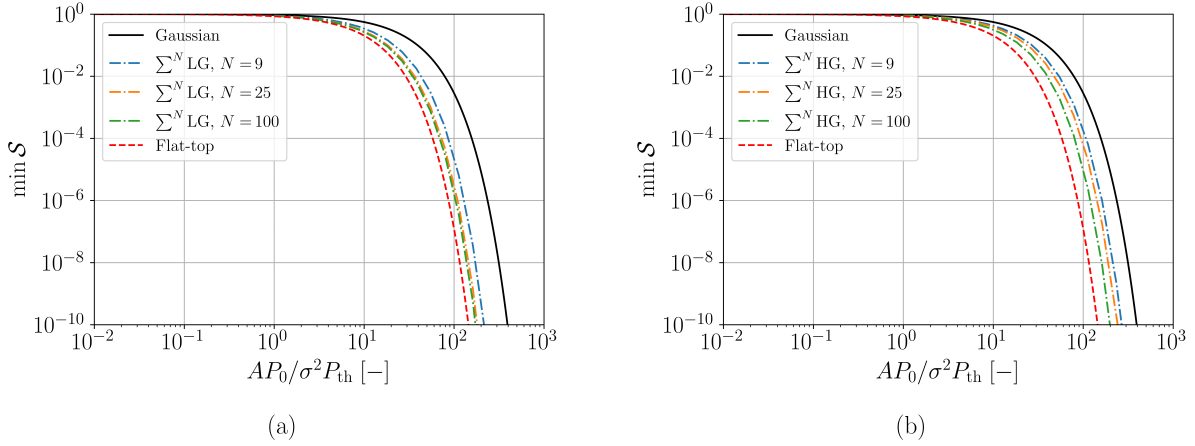}
    \caption{Performance parameter $\min\mathcal S$ for an incoherent superposition of (a) Laguerre-Gaussian and (b) Hermite-Gaussian beams, for different numbers of modes $N$.}
    \label{fig_incoh_smin}
\end{figure}

Using the method above, the approximated super-Gaussian beams of different orders can be obtained using both Laguerre-Gaussian and Hermite-Gaussian modes. In Fig.~\ref{fig_incoh_fields} the best fits to a super-Gaussian beam of 8th order are shown for both truncated LG and HG bases. By increasing the number of modes considered, higher-order super-Gaussian beams can be targeted, and the root mean square error (RMSE) of the resulting irradiance can be decreased, as seen in Fig.~\ref{fig_incoh_rmse}. It can be seen in this figure that the LG modes converge faster than the HG modes towards super-Gaussian beams. Furthermore, the results also show that a higher number of modes are required as the order of the super-Gaussian beam increases, as expected. 
To evaluate the communication performance of these beams, the probability $\min\mathcal S$ can be computed for different $AP_0/\sigma^2P_\mathrm{th}$ values for the beam shapes obtained through incoherent superposition. The results for different numbers of modes along with the baseline Gaussian solution and the optimum flat-top are shown in Fig.~\ref{fig_incoh_smin}. It can be seen that the performance of the incoherent superposition is better than the baseline Gaussian beam, and that the performance of LG beams is better than the HG beams, as expected from the better fit to the super-Gaussian beams discussed above.

\subsection{Coherent light: Phase mask}\label{sec_coherent}
\begin{figure}[!htb]
    \centering
    \includegraphics[width=0.8\linewidth]{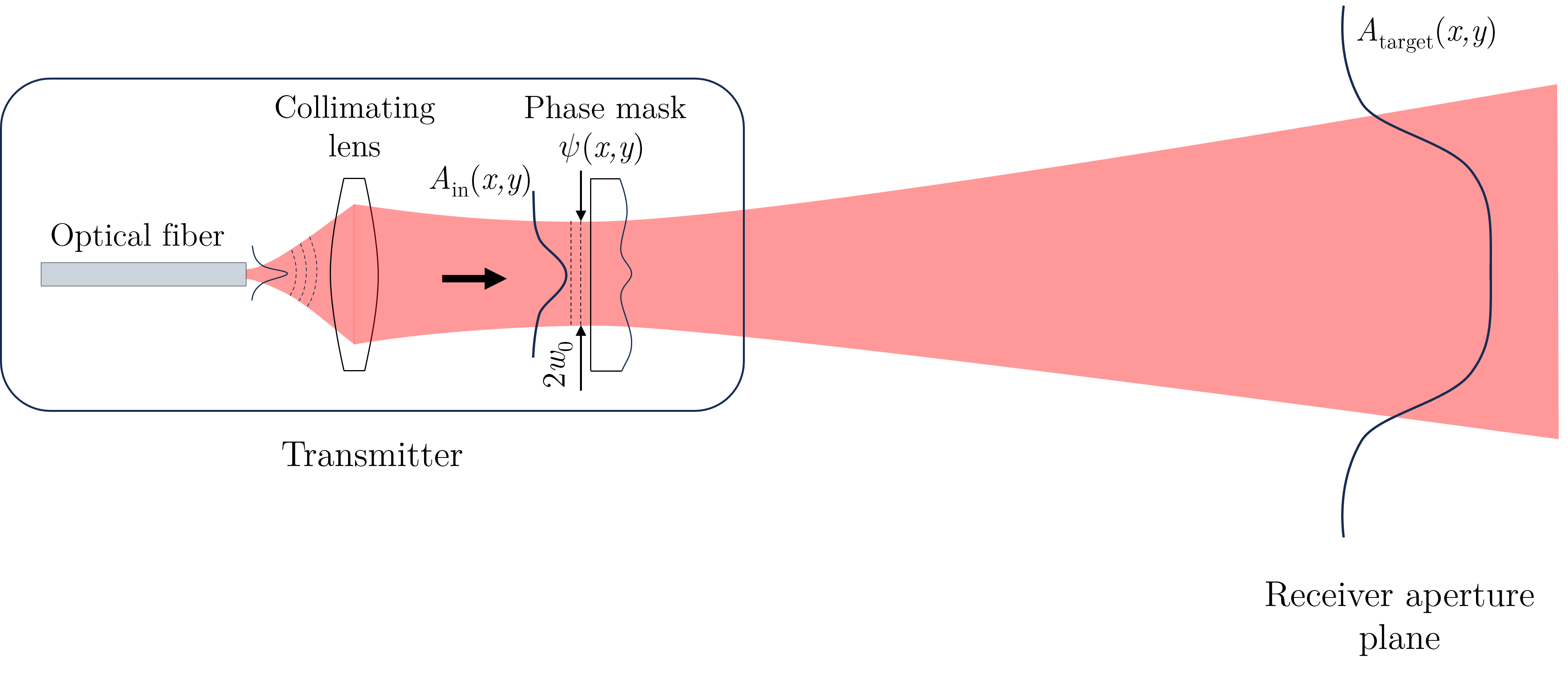}
    \caption{Illustration of the coherent beam shaping technique using a phase mask.}
    \label{fig_pm_illustration}
\end{figure}

The previous method can be computationally too expensive, and the truncation order of the basis also limits the obtainable beam shape to a restricted domain. Furthermore, using the polarization and wavelength degrees of freedom to incoherently combine the modes limits the applicability of these degrees of freedom for modulation, multiplexing, and receive/transmit optical path splitting. To overcome these limitations, a coherent beam shaping technique can be followed. By considering a fully coherent beam of light in the source, a properly designed phase mask can be used to generate different beam shapes (see Fig.~\ref{fig_pm_illustration}). For optical communication transmitter terminals, the beam from the source can be considered to be Gaussian, as most of the terminals are built from single-mode fiber components \cite{badas_optimum_2024}.

There are several optical components that can induce the bidimensional phase masks\footnote{In general, beam shaping can be achieved using a combined phase-transmission mask. However, in the context of this work, the power loss introduced by transmission masks is detrimental to system performance. Therefore, only phase-only masks are considered.} needed for coherent beam shaping. For mitigation of atmospheric turbulence, Billaud et al. proposed the use of multi-plane light converters for coherent beam shaping \cite{billaud_10_2023}. Multi-plane light converters consist of both reflective and refractive optical phase masks that convert different spatial modes into one another \cite{zhang_multi-plane_2023}. Another way of generating a beam shape of interest in the far-field is through the use of metasurfaces, in which subwavelength geometries are used to manipulate the phase induced at each point of the surface (although many other functionalities can also be achieved with these structures) \cite{badas_metalens_2025}. Dynamical optical elements such as spatial light modulators, or deformable mirrors, are used for applications in which the phase mask to be applied changes over time \cite{wang_variable_2012, burger_implementation_2014, li_using_2018, hayasaki_automatic_2023}. For the static beam shaping application targeted in this work, diffractive optical elements seem to be one of the most suitable experimental implementations for our phase screens \cite{xu_fabrication-integrated_2025}. This is because they can be fabricated with high precision \cite{xu_fabrication-integrated_2025, buske_diffractive_2025}, introduce well-defined phase modulations with high optical transmission efficiency, and operate as compact passive components that do not require active control during operation. The performance of such systems has been demonstrated in Refs.~\cite{buske_advanced_2022, buske_diffractive_2025}, where multiple cascaded diffractive optical elements --following a concept similar to multi-plane light converters-- are used for laser material processing, enabling the generation of flat-top and other tailored beam shapes. In Ref.~\cite{buske_diffractive_2025}, the effects of the finite steepness of the $2\pi$ phase jumps arising from manufacturing limitations are analyzed. Although binary optical phase elements have been proposed for flat-top beam shaping \cite{yang_analysis_2003}, they are generally less flexible and typically exhibit reduced beam quality compared to multi-level or continuously varying phase profiles.


Considering the source of light to be Gaussian at the transmitter end, a phase mask can be designed that converts this beam into a super-Gaussian–like irradiance distribution in the far-field (see Fig.~\ref{fig_pm_illustration}). To achieve this, a phase mask can be designed employing phase retrieval algorithms \cite{taylor_phase_1981, fienup_phase_1982, fienup_phase_2013, shechtman_phase_2015, guo_review_2017, ripoll_review_2004}. These algorithms --one of the most common being the Gerchberg–Saxton (GS) algorithm or modified versions of it-- employ an iterative process that optimizes the targeted far-field irradiance. The basic principle is that the algorithm alternates between the spatial domain (phase mask plane) and the Fourier domain (far-field) using forward and inverse Fourier transforms. In each domain, the amplitude constraint is enforced (Gaussian amplitude at the transmitter plane and the desired super-Gaussian amplitude in the far-field), while the phase is allowed to evolve during the iterations. Starting from an initial guess for the phase $\psi_0(x,y)$, the complex field
\(
U_0(x,y) = A_{\mathrm{in}}(x,y)e^{i\psi_0(x,y)}
\)
is propagated to the far-field through a Fourier transform
\(
\tilde{U}_0(k_x,k_y) = \mathcal{F}\left\{{U_0(x,y)}\right\}.
\)
The amplitude is then replaced by the desired far-field amplitude $A_{\mathrm{target}}(k_x,k_y)$ while keeping the phase, and the field is propagated back with an inverse Fourier transform. Repeating this process iteratively yields a phase distribution that produces approximately the desired far-field irradiance \cite{ripoll_review_2004}. Moreover, several modifications of the Gerchberg–Saxton algorithm have been proposed to improve convergence speed and generate continuous phase screens \cite{dixit_designing_1996, lin_distributed_1995, deng_pure-phase_1996, marozas_fourier_2007, dixit_kinoform_1994}. For example, a modified iterative scheme presented in Ref.~\cite{liu_iterative_2002} demonstrates the generation of a far-field irradiance that approximates to a super-Gaussian beam of 12th order. Other ways of designing phase masks have also been proposed, such as the Zernike basis decomposition combined with grid-search and gradient-descent optimization introduced in Ref.~\cite{badas_metalens_2025}. This method exploits the azimuthal symmetry of the problem, yielding radially symmetric phase masks. However, despite this advantage, iterative phase retrieval algorithms such as Gerchberg–Saxton generally exhibit faster convergence and remain among the most widely used approaches for practical phase mask design. Furthermore, recent advances in the training of neural networks can also be exploited to design these phase masks iteratively, exploiting the concept of diffractive neural networks \cite{buske_advanced_2022}.

\begin{figure}[!htb]
    \centering
    \includegraphics[width=0.9\linewidth]{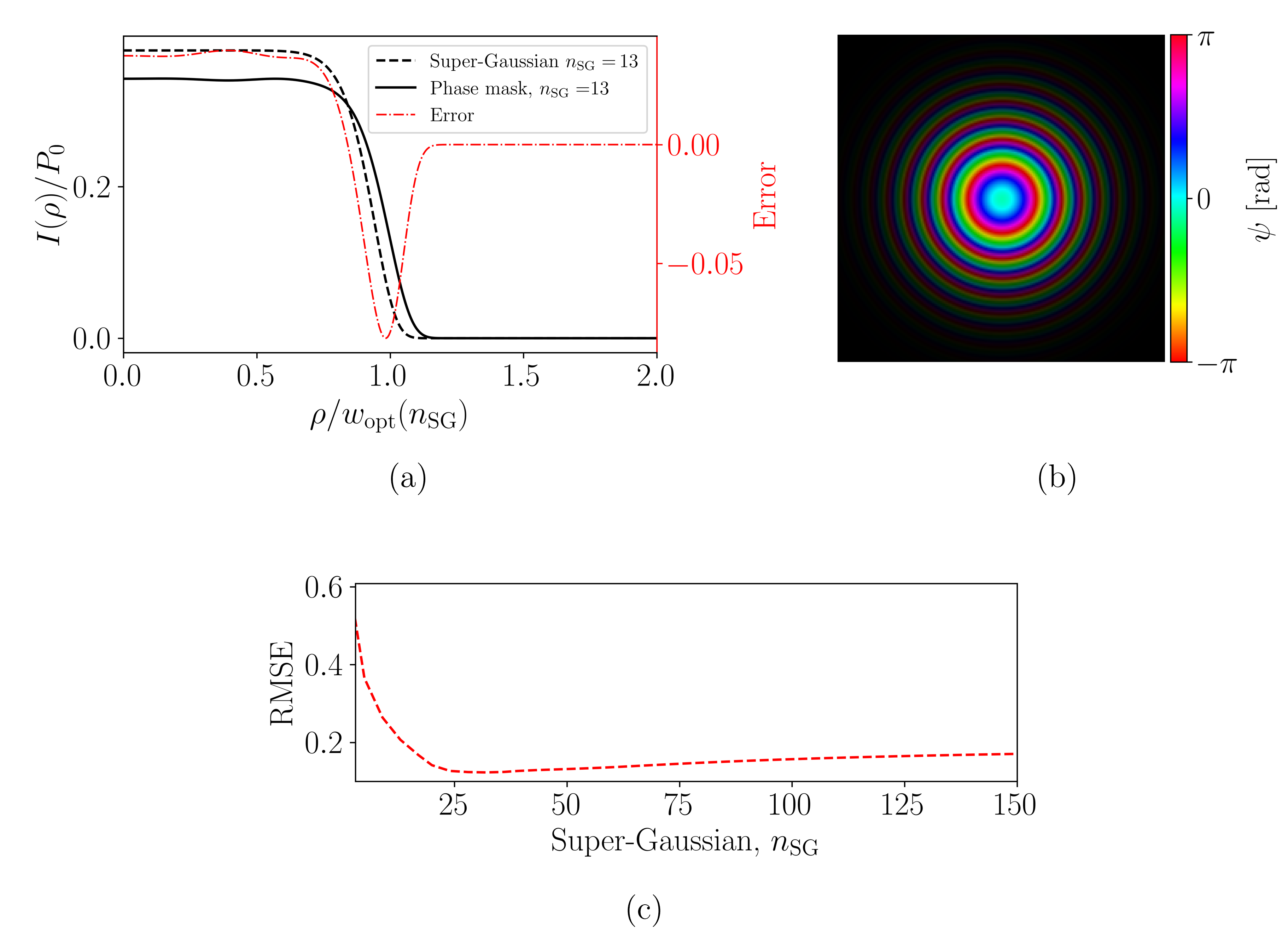}
    \caption{Coherent beam shaping using the GS algorithm for obtaining super-Gaussian beams. (a) shows the targeted 13th order super-Gaussian and the achieved far-field irradiance, and (b) shows the input beam, where the irradiance is Gaussian (brightness of the color) and the retrieved phase mask is given by the colormap. (c) shows the root mean square error of the far-field irradiance compared to the targeted irradiance for increased super-Gaussian orders targeted.}
    \label{fig_coh_fields_rmse}
\end{figure}

Using the Gerchberg-Saxton algorithm proposed by Liu et al. \cite{liu_iterative_2002}, the phase masks have been designed to convert the Gaussian beam into different super-Gaussian orders. To do so, the algorithm is initialized using a phase mask obtained relying solely on geometrical optics principles \cite{bryngdahl_geometrical_1974}. After one hundred iterations, the retrieved phase screen is obtained and the resulting far-field irradiance is analyzed. As an example, the targeted far-field super-Gaussian beam and the achieved approximated beam through this method are shown in Fig.~\ref{fig_coh_fields_rmse}(a)-(b). The irradiances shown are normalized to total unit power. Furthermore, the retrieved phase mask obtained is also shown. Due to the azimuthally symmetrical nature of the problem and the azimuthally symmetrical initialization of the phase mask provided by the geometrical optical phase mask, the numerically retrieved phase mask also possesses this symmetry. In case random phase masks would have been used \cite{dixit_designing_1996}, this symmetry would be broken, and some more steps would need to be included in the algorithm to force the phase mask to have this symmetry. 

In Fig.~\ref{fig_coh_fields_rmse}(c), the root mean square error of the obtained far-field is compared with the targeted super-Gaussian for different orders. For approximating the Gaussian beam $n_\mathrm{SG}=2$ the GS algorithm is not too appropriate, as it can be seen. This might be due to the initialization of the phase mask. From a certain point on ($n_\mathrm{SG}\approx25$), as the targeted super-Gaussian order increases, the algorithm starts to deviate further from the target as it can not accurately approximate the discontinuity generated in the $n_\mathrm{SG}\to\infty$ flat-top limit. 

\begin{figure}
    \centering
    \includegraphics[width=0.5\linewidth]{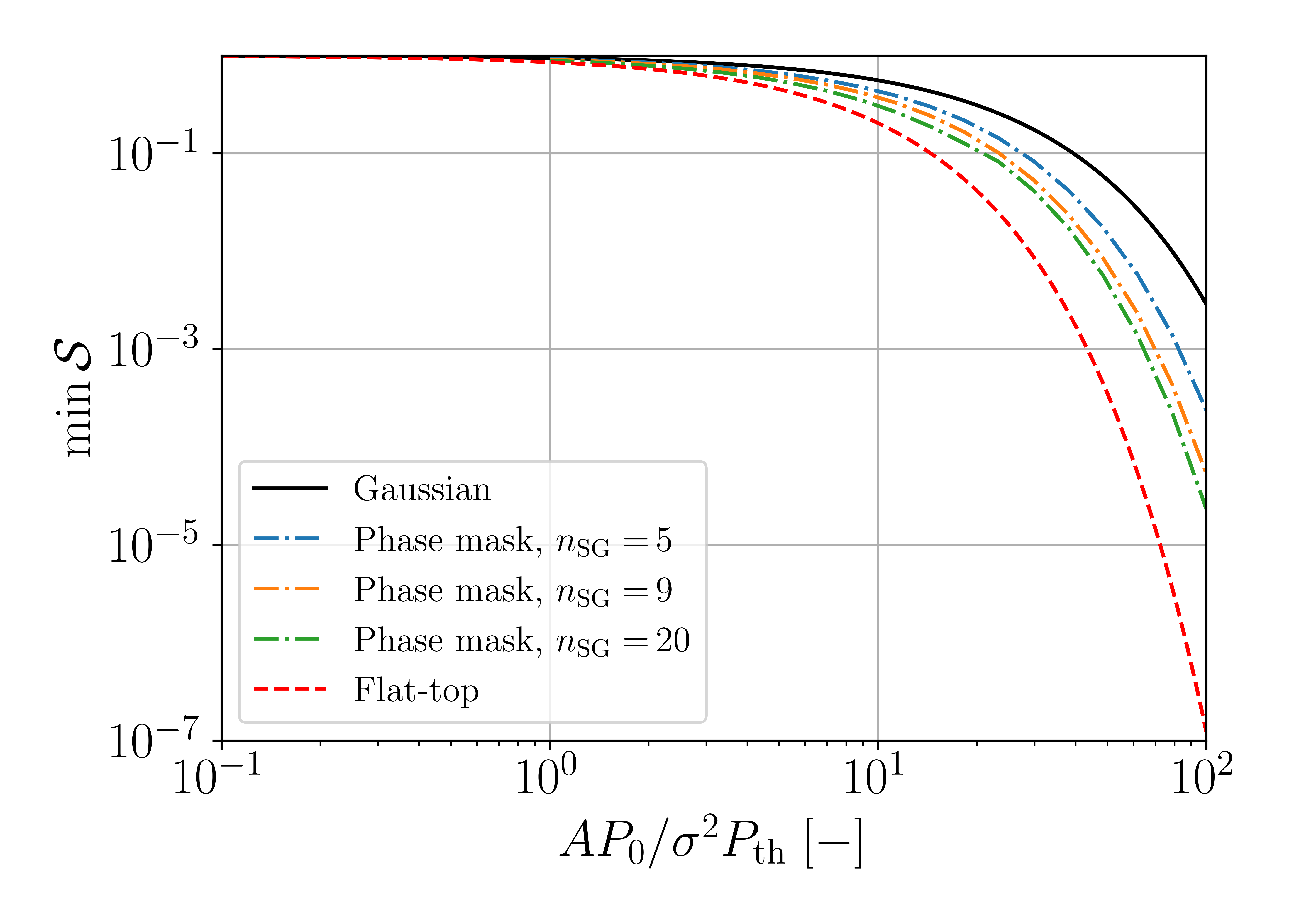}
    \caption{Performance parameter $\min\mathcal S$ for the irradiance obtained with the phase screens computed through the GS algorithm for different super-Gaussian beam orders, compared to the conventional Gaussian and the optimum flat-top performances.}
    \label{fig_coh_smin}
\end{figure}

In a similar manner to what was done for the incoherent case, the performance parameter $\min\mathcal S$ for far-field irradiances obtained through the designed phase masks has been evaluated. The results for different targeted super-Gaussian beams are shown in Fig.~\ref{fig_coh_smin}. The results show a close resemblace to the analytically obtained results in Fig.~\ref{fig_G_SG_FT}(b). However, as the super-Gaussian order increases, the far-field irradiances obtained through the optimized phase mask can no longer closely resmeble the respective super-Gaussian beam. This produces the performance saturation of these physically achievable beams, to occur faster than the analytical ones (that saturate at the flat-top curve as the super-Gaussian order increases as shown in Fig.~\ref{fig_G_SG_FT}).

\begin{figure}[!htb]
    \centering
    \includegraphics[width=0.5\linewidth]{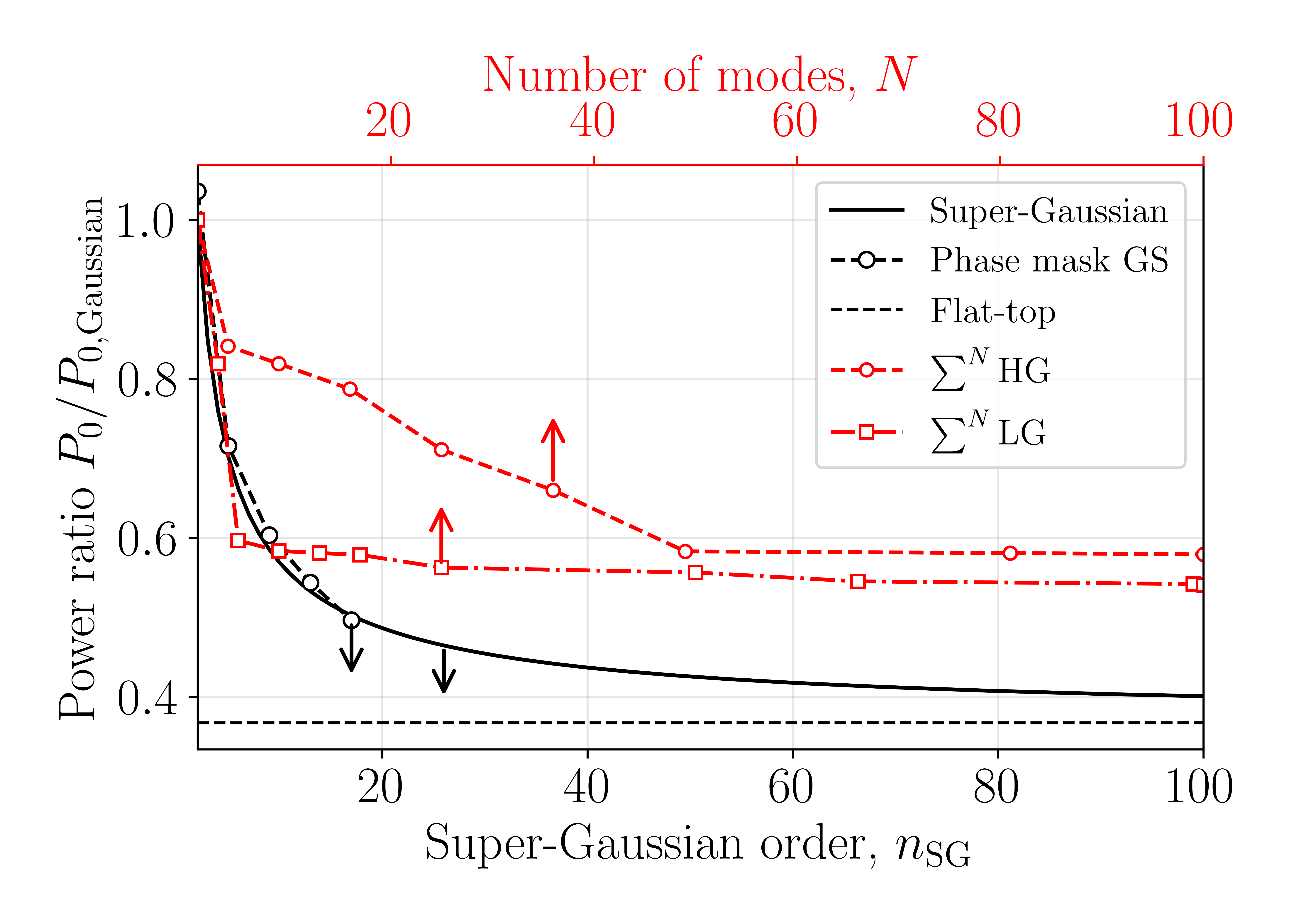}
    \caption{Power savings (compared to a conventional Gaussian beam) of the theoretical super-Gaussian, flat-top, incoherent beamshaping with LG and HG beam families, and coherent beam shaping achieved through a phase mask.}
    \label{fig_coh_incoh_gain}
\end{figure}

For practical purposes, it might be of interest to the reader to compare the performance of both the theoretical super-Gaussians, and the approximated incoherent and coherent beams. To do so, the ratio of the transmitted power $P_0$ required to obtain the same performance $\min\mathcal{S}$ as that of a conventional Gaussian beam is computed. The results are shown in Fig.~\ref{fig_coh_incoh_gain} where the ratio $P_0/P_{0,\mathrm{Gaussian}}$ is plotted as a function of the super-Gaussian order or the number of modes considered. For the super-Gaussian and flat-top beams, the curve is identical to that shown in Fig.~\ref{fig_gain_n_SG_FT}, and it is seen that the super-Gaussian asymptotically approaches the flat-top behaviour as the order $n_\mathrm{SG}$ increases. For the incoherent superposition of LG and HG families presented in \S~\ref{sec_incoherent}, the plot shows the power ratio as the number of modes considered in the basis is increased. Following the results shown in Figs.~\ref{fig_incoh_rmse}-\ref{fig_incoh_smin}, the LG beams converge faster than the HG and saturate at around $\sim53\%$ (although the ratio continues to decrease slowly for higher $N$ numbers). As discussed in \S~\ref{sec_incoherent}, polarization based-superposition would be limited to $N=2$ modes. For the LG beams families, the obtained $\sim60\%$ ratio is in agreement with the power saving results reported in Ref.~\cite{badas_optimum_2024}.


For the coherently generated beam shapes, the phase masks obtained through the GS algorithm provide results that are extremely close to the theoretical super-Gaussian ratios up to approximately the 17th super-Gaussian order. However, for higher orders, as shown in Fig.~\ref{fig_coh_fields_rmse}(c), the phase masks designed with this approach begin to diverge from the corresponding super-Gaussian profiles. For $n_\mathrm{SG} > 17$, the phase masks intended to generate higher-order super-Gaussian beam shapes instead produce beams that perform worse than those designed for lower orders. For the phase mask designed for $n_\mathrm{SG} = 17$, the resulting far-field irradiance yields a performance ratio of $50\%$, in agreement with the analytical result for the targeted super-Gaussian profile. This ratio can also be interpreted as the required value of the parameter $AP_0/\sigma^2 P_\mathrm{th}$ relative to that of a conventional Gaussian beam. Consequently, the graph may also be interpreted as indicating the reduction in required receiver aperture area for achieving the same performance, for example.

It is important to note that, when evaluating the performance of both the incoherent and coherent beam shapes, the far-field irradiance beam width $w_\mathrm{opt}$ had to be scaled. In our analysis, we did not consider limitations associated with adjusting the far-field beam width due to the finite aperture of the transmitter, nor the truncation effects on the transmitted beams \cite{badas_seidel_2025}. As the targeted far-field beam width decreases—corresponding to the small $AP_0/P_\mathrm{th}$ regime shown in Fig.~\ref{fig_G_SG_FT}(a)—the divergence of the beams involved may limit the minimum achievable beam width in the far-field. This limitation can be mitigated by increasing the transmitter aperture or by partially sacrificing the minimum achievable performance $\mathcal{S}$.

\section{Conclusions and future work}
This study investigated the optimum far-field irradiance for mitigating the effect of the transmitter pointing jitter in intersatellite optical communication.
It has been found through a variational approach that the optimum far-field irradiance corresponds to a flat-top beam in the far-field. Due to the physics of wave propagation limiting the obtention of a flat-top in the far-field, the performance of the super-Gaussian beam family has been investigated (as it approaches the flat-top beam in the infinite order limit). It has been found theoretically that a flat-top requires $e^{-1}\approx36,8\%$ of the power required by the conventional Gaussian beam to achieve the same performance. Furthermore, incoherent and coherent beam shaping techniques have been proposed to approximate super-Gaussian beams of different orders in the far-field. It has been found that incoherent beam shaping using Laguerre-Gaussian beams converges faster to the super-Gaussian beams than the Hermite-Gaussians due to the azimuthal symmetry of the former. The achievable minimum power ratio through incoherent beam shaping was found to be $\sim53\%$ when combining one hundred Laguerre-Gaussian modes. Coherent beam shaping, which involved designing a phase mask through the Gerchberg-Saxton algorithm, was able to achieve a very close result to that given by a theoretical super-Gaussian beam of 17th order. In this case, yielding to a power ratio of $\sim50\%$ compared to a conventional Gaussian beam.

These results extend the previous attempt to optimize the performance of intersatellite optical communication affected by transmitter pointing jitter \cite{toyoshima_optimum_2002, badas_optimum_2024} by providing a fundamental lower limit and approximating it via different optical beam shaping methods.
It is important to mention that while the optimization provided is done for minimizing probabilities of being below a reasonable power thershold, other performance parameters (e.g., average capacity, peak capacity) would yield different optimum far-field irradiances. 

Future work could further expand the beam shaping approaches explored in this work in several directions. First, improved algorithms for designing phase masks capable of generating higher-order super-Gaussian beams would allow more optimum far-field irradiances. A realistic performance assessment should also incorporate losses that occur in a real system, similar to the analysis presented in \cite{badas_annular_2026}, in order to evaluate the actual power savings of the proposed beam shapes. In addition, experimental implementations of the incoherent superposition of higher-order modes—such as through multi-plane light converters or comparable optical architectures—would provide a pathway toward practical realization of the concepts discussed in \S~\ref{sec_incoherent}. Experimental demonstrations of the coherent beam shaping techniques are also required, particularly accounting for manufacturing constraints associated with phase mask elements such as diffractive optical elements. Furthermore, a detailed study of the impact of optical aberrations of the incoming Gaussian beam in coherent beam shaping systems would help quantify performance degradation in realistic setups. Finally, extending the proposed beam shaping to relevant free space optical communication scenarios would be valuable, including coupling efficiency into detectors or optical fibers in satellite optical communication systems affected by receiver pointing jitter, as well as downlink scenarios with atmospheric turbulence.




\vspace{1cm}
\noindent\textbf{Acknowledgments.} M.B. would like to thank the Nederlandse Organisatie voor Wetenschappelijk Onderzoek for funding this research (P19-13).

\bibliography{references}
\bibliographystyle{ieeetr}

\appendix
\clearpage
\section{Proofs for \S \ref{sec_variational_FP}}\label{app_proofs}
{\textbf{Proof 1.}}
Here is a simple argument showing that you cannot have a collected finite power at any displacement $(x_0,y_0)$ of the beam in the real bidimensional plane 
\[
g(x_{0},y_{0})
=\iint_{\mathbb{R}^{2}}I(x,y)\;\mathcal{A}(x-x_{0},\,y-y_{0})\,dx\,dy
\;\ge\;P_{\rm th}
\quad\forall\,(x_{0},y_{0})\in\mathbb{R}^2
\]  
while at the same time having a finite total power of the beam
\[
P_{0}
=\iint_{\mathbb{R}^{2}}I(x,y)\,dx\,dy
\;<\;\infty.
\]
\begin{enumerate}
    \item Partition the plane into infinitely many disjoint translates of the annular aperture.  
       Let \(\{(x_{0}^{(k)},y_{0}^{(k)})\}_{k=1}^{\infty}\) be a collection of centers so that the sets  
       \[
         \mathcal{A}_{k}\;=\;\{(x,y)\,:\,a< (x-x_{0}^{(k)})^{2}+(y-y_{0}^{(k)})^{2}<b\}
       \]
       are pairwise disjoint.  (You can do this because the plane admits a packing of disjoint congruent annuli, the argument is also valid for other kinds of aperture shapes)
    \item By hypothesis each translate carries at least \(P_{\rm th}\) power  
       \[
       \iint_{\mathcal{A}_{k}}I(x,y)\,dx\,dy
       \;=\;
       g\bigl(x_{0}^{(k)},y_{0}^{(k)}\bigr)
       \;\ge\;P_{\rm th}.
       \]
    \item Since the \(\mathcal{A}_{k}\) are disjoint, the total power is at least the sum over all \(k\)  
       \[
       P_{0}
       =\iint_{\mathbb{R}^{2}}I(x,y)\,dx\,dy
       \;\ge\;
       \sum_{k=1}^{\infty}
         \iint_{\mathcal{A}_{k}}I(x,y)\,dx\,dy
       \;\ge\;
       \sum_{k=1}^{\infty}P_{\rm th}.
       \]
       But \(\sum_{k=1}^{\infty}P_{\rm th}=\infty\) (since \(P_{\rm th}>0\)), so this forces   
       \[
         P_{0}=\infty,
       \]  
       contradicting the finiteness assumption \(P_0<\infty\).
\end{enumerate}
Hence, it is impossible to have  
\[
g(x_{0},y_{0})\ge P_{\rm th}>0
\quad\text{for all }\;(x_{0},y_{0}),
\]  
because that would force the total beam power \(P_{0}\) to be infinite.

\vspace{0.5cm}
\textbf{Proof 2}. Here we wish to proof that the bidimensional integral of the function $g(x_0,y_0)$ defined in Eq.~\eqref{eq_gx0y0} is finite,
\[
G \;=\;\iint_{\mathbb{R}^2}g(x_0,y_0)\,dx_0\,dy_0
<\infty.
\]
\begin{enumerate}
    \item Write \(G\) as an iterated integral
    \[
    G
    =\iint_{\mathbb{R}^2}\Biggl[\iint_{\mathbb{R}^2}
    I(x,y)\,{\mathcal A}(x-x_0,y-y_0)\,dx\,dy\Biggr]
    \,dx_0\,dy_0.
    \]
    \item By Fubini (all functions are nonnegative), swap the order of integration
    \[
    G
    =\iint_{\mathbb{R}^2}I(x,y)
    \Biggl[\iint_{\mathbb{R}^2}{\mathcal A}(x-x_0,y-y_0)\,dx_0\,dy_0\Biggr]
    \,dx\,dy.
    \]
    \item Observe that
    \[
    \iint_{\mathbb{R}^2}{\mathcal A}(x-x_0,y-y_0)\,dx_0\,dy_0
    \;=\;
    \iint_{\mathbb{R}^2}{\mathcal A}(u,v)\,du\,dv
    \;=\;
    \bigl|\mathcal A\bigr|
    \;<\;\infty,
    \]
    where we have changed variables \(u=x-x_0\), \(v=y-y_0\).  By hypothesis \(\bigl|\mathcal A\bigr|\) (the area of the aperture) is finite.
    \item Hence
    \[
    G
    =\iint_{\mathbb{R}^2}I(x,y)\,\bigl|\mathcal A\bigr|\;dx\,dy
    =\bigl|\mathcal A\bigr|\;\iint_{\mathbb{R}^2}I(x,y)\,dx\,dy
    =\bigl|\mathcal A\bigr|\;P_0
    \;<\;\infty.
    \]
\end{enumerate}
Since \(0<P_0<\infty\) and \(0<\lvert\mathcal A\rvert<\infty\), it follows that \(0<G<\infty\). 

\vspace{0.5cm}
\textbf{Proof 3}. Here we wish to proof that among all nonnegative \(g(x_0,y_0)\) with fixed total mass 
\[
\iint_{\mathbb{R}^2}g(x_0,y_0)\,dx_0\,dy_0 = G,
\]
the one that maximizes the area
\[
A = \bigl|\{(x_0,y_0):g(x_0,y_0) > P_{\rm th}\}\big|
\]
is a circular flat-top function of height \(P_{\rm th}\).  Moreover, the maximum area that can be covered is a circular area $A_{\max}$ with radius $r_0$
\[
A_{\max} \;=\;\frac{G}{P_{\rm th}}
\quad\Longrightarrow\quad
\text{radius }r_0 = \sqrt{\frac{A_{\max}}{\pi}}
=\sqrt{\frac{G}{\pi\,P_{\rm th}}}\,. 
\]
\begin{enumerate}
    \item For any nonnegative measurable \(g\) one has the identity
    \[
    \iint_{\mathbb{R}^2}g(x,y)\,dx\,dy
    \;=\;\int_{0}^{\infty}
    \Bigl|\{(x,y):\,g(x,y)>t\}\Bigr|\;dt.
    \]
    In particular, split the integral at \(t=P_{\rm th}\):
    \[
    G
    =\int_{0}^{P_{\rm th}}
    \bigl|\{g>t\}\bigr|\,dt
    \;+\;\int_{P_{\rm th}}^{\infty}
    \bigl|\{g>t\}\bigr|\,dt.
    \]
    Since \(\{g>t\}\supseteq\{g>P_{\rm th}\}\) for all \(t<P_{\rm th}\), we get a lower bound
    \[
    \int_{0}^{P_{\rm th}}
    \bigl|\{g>t\}\bigr|\,dt
    \;\ge\;
    \int_{0}^{P_{\rm th}}A\,dt
    \;=\;P_{\rm th}\,A,
    \]
    where \(A:=|\{g>P_{\rm th}\}|\).  Also the second integral is nonnegative.  Hence
    \[
    G \;\ge\;P_{\rm th}\,A
    \;\Longrightarrow\;
    A \;\le\;\frac{G}{P_{\rm th}}.
    \]
    So no function can have \(\bigl|\{g>P_{\rm th}\}\bigr|\) larger than \(G/P_{\rm th}\).
    \item Achieving the bound by a $\text{circ}$ function. To see that the bound is attained, simply take
    \begin{equation}
    g(x,y)=
    \begin{cases}
    P_{\rm th},&x^2+y^2\le r_0^2,\\
    0,&\text{otherwise},
    \end{cases}
    \end{equation}
    and choose \(r_0\) so that its total mass is \(G\):
    \[
    \iint_{\mathbb{R}^2}g\,dx\,dy
    =\iint_{x^2+y^2\le r_0^2}P_{\rm th}\,dx\,dy
    =P_{\rm th}\,\bigl(\pi\,r_0^2\bigr)
    \stackrel!=G
    \quad\Longrightarrow\quad
    r_0^2=\frac{G}{\pi\,P_{\rm th}}.
    \]
    Then \(\{g>P_{\rm th}\}\) is exactly the disk of radius \(r_0\), whose area is
    \[
    \pi\,r_0^2
    =\pi\;\frac{G}{\pi\,P_{\rm th}}
    =\frac{G}{P_{\rm th}}.
    \]
\end{enumerate}

\vspace{0.5cm}
\textbf{Proof 4}. Below is a direct way to see that there is no nonnegative, compactly supported function \(I(x,y)\) whose finite‐aperture convolution
\[
g(x_0,y_0)\;=\;\iint_{\mathbb{R}^2}I(x,y)\,\mathcal{A}\bigl(x-x_0,y-y_0\bigr)\,dx\,dy
\]
is exactly the sharp top‐hat of radius \(r_0\) and height \(P_{\rm th}\). Furhtermore, the exact solution in the small aperture approximation for the farfield irradiance $I$ is obtained.
\begin{enumerate}
    \item  Let \(\mathcal{A}(u,v)\) be aperture function of a finite aperture, say a disk of radius \(a\)
   \[
   \mathcal{A}(u,v)=
   \begin{cases}
     1,&u^2+v^2\le a^2,\\
     0,&\text{otherwise}.
   \end{cases}
   \]
   Then \(g\) is the two‐dimensional convolution of \(I\) with \(\mathcal{A}\).
    \item In the Fourier domain, convolution becomes multiplication.  Denoting by \(\tilde{I}(k_x,k_y)\), \(\tilde{\mathcal{A}}(k_x,k_y)\), \(\tilde{g}(k_x,k_y)\)
   the Fourier transforms,
   \[
   \tilde{g}(k_x,k_y)
   \;=\;
   \tilde{I}(k_x,k_y)\;\tilde{\mathcal{A}}(k_x,k_y)\,.
   \]
   Hence any candidate \(I\) must satisfy
   \[
   \tilde{I}(k_x,k_y)
   \;=\;
   \frac{\tilde{g}(k_x,k_y)}{\tilde{\mathcal{A}}(k_x,k_y)}\,.
   \]
    \item But \(\tilde{g}\) is the transform of a disk of radius \(r_0\) and height
   \(P_{\rm th}\):
   \[
   g(x_0,y_0)=
   \begin{cases}
     P_{\rm th},&x_0^2+y_0^2\le r_0^2,\\
     0,&\text{otherwise},
   \end{cases}
   \]
   so
   \[
   \tilde{g}(k)
   =P_{\rm th}\,\pi\,r_0^2\;\frac{2\mathcal{J}_1\bigl(|k|\,r_0\bigr)}{|k|\,r_0},
     \quad|k|=\sqrt{k_x^2+k_y^2}.
   \]
   Likewise \(\tilde{\mathcal{A}}(k)=\pi\,a^2\,(2\mathcal{J}_1(|k|a)/(|k|a))\).  Therefore
   \begin{equation}\label{eq_I(k)}
   \tilde{I}(k)
   =\;P_{\rm th}\;\frac{r_0}{a}
   \;\frac{\mathcal J_1(|k|\,r_0)}{\mathcal J_1(|k|\,a)}.
   \end{equation}
    \item The inverse Fourier transform (the Hankel transform of the first kind) of this ratio is a function in polar coordinates given as
    \begin{equation}
        I(\rho) = \int_0^\infty\tilde{I}(k)\;\mathcal J_1(kr) \;\rho\;d\rho
    \end{equation}
    which in fact blows up and changes sign (because of the zeros of \(\mathcal J_1\)), so one cannot get a strictly nonnegative, finite‐supported \(I\ge0\).  
    \item However, for small aperture sizes $a\to0$ we can expand the $\tilde{\mathcal{A}}(k)$ as
    \begin{equation}
        \lim_{a\to0}\tilde{\mathcal{A}}(k)=\dfrac{a^2}{2}+\mathcal{O}(a^4)
    \end{equation}
    Hence, $\tilde{I}(k)$ can be written as
    \begin{equation}
        \lim_{a\to0}\tilde{I}(k)=\dfrac{2P_{\rm th}r_0}{a^2}
   \;J_1(|k|\,r_0).
    \end{equation}
    Yielding the inverse Hankel transform
    \begin{equation}
        I(\rho)=\begin{cases}
            \dfrac{2P_\text{th}r_0}{a^2} &\rho\leq r_0\\
            0 &\text{otherwise}
        \end{cases}
    \end{equation}
    which is the flat-top beam. The normalization of the obtained flat-top might not be correct due to the $a\to0$ approximation. To correctly normalize, the total power normalization condition can be included.
\end{enumerate}
Hence, there is no nonnegative function \(I\) of finite support whose convolution with a finite aperture \(\mathcal{A}\) is exactly the sharp flat-top function of radius \(r_0\).  Equivalently, the deconvolution kernel \(\tilde{g}/\tilde{\mathcal{A}}\) has poles (zeros of \(\tilde{\mathcal{A}}\)) and forces sign‐changes.
In practice, one can only approximate the step‐function \(g\) by choosing \(\tilde{I} = \tilde{g}/\tilde{\mathcal{A}}\) through numerical evaluation.

\section{Ring flat-top beam}\label{app_ring}
The reader could be wondering if a ring type flat-top of the kind
\begin{equation*}
    I(\rho) = \begin{cases}
    \dfrac{P_\text{th}}{A},\quad &r_1<\rho<r_2 \\
    0, \quad &\text{else}
    \end{cases}
\end{equation*}
would be better than the flat-top, where the $r_1$ and $r_2$ are the inner and outer radius of the ring and the power of the beam is $P_0$. To show that in this case the optimum is found when $r_1=0$, hence a flat-top beam, we obtain the value of $\mathcal{S}$ for this shape in a similar manner to what was done for the flat-top beam
\begin{equation*}
    \mathcal{S}=\int_{0}^{r_1}  f_\mathrm{R}(\rho)  \;d\rho+\int_{r_2}^\infty  f_\mathrm{R}(\rho)  \;d\rho
\end{equation*}
that under the condition that the power of the beam is $P_0$ we have
\begin{equation*}
    \pi(r_2^2-r_1^2)\dfrac{{P_\text{th}}}{A}=P_0\xrightarrow[]{}r_2=\sqrt{\dfrac{P_0}{\pi {P_\text{th}}}+r_1^2}
\end{equation*}
and then the performance can be written as a function of $r_1$ as
\begin{equation*}
\begin{split}
    \mathcal{S}(r_1)&=\int_{0}^{r_1}  f_\mathrm{R}(\rho)  \;d\rho+\int_{\sqrt{\frac{P_0}{\pi {P_\text{th}}}+r_1^2}}^\infty  f_\mathrm{R}(\rho)  \;d\rho\\
    &=\dfrac{1}{2\pi}\left[1+e^{-\frac{r_1^2}{2\sigma^2}}\left(e^{-\frac{P_0}{2\pi P_\text{th} \sigma^2}}-1\right)\right]
    \end{split}
\end{equation*}
Now we have to find the $r_1$ at which $\mathcal{S}$ is minimum so
\begin{equation*}
\begin{split}
\dfrac{d\mathcal{S}(r_1)}{dr_1}&=0 \\
&=-\dfrac{r_1}{2\pi \sigma^2}\left[e^{-\frac{r_1^2}{2\sigma^2}}\left(e^{-\frac{P_0}{2\pi P_\text{th} \sigma^2}}-1\right)\right]
    \end{split}
\end{equation*}
which has solutions at $r_1=0$ and $r_1\to\infty$. It can be seen that the extrema at $r_1$ is a minimum by the second derivative test 
\begin{equation*}
    \dfrac{d^2\mathcal{S}(r_1)}{dr_1^2}\Bigg|_{r_1=0} = \dfrac{1-e^{-\frac{P_0}{2\pi P_\text{th} \sigma^2}}}{2\pi\sigma^2}>0
\end{equation*}
because $P_0/(2\pi P_\text{th} \sigma^2)>0$ and $\sigma^2>0$. Hence, we see that the ring flat-top of the type analysed here, converges to the flat-top at its maximum performance.

\clearpage

\end{document}